\documentclass[numbers,preprint,12pt]{elsarticle}

\usepackage[margin=2.2cm]{geometry}
\usepackage{framed,multirow}

\usepackage{amssymb}
\usepackage{latexsym}

\usepackage{hyperref,url}
\hypersetup{colorlinks=true, urlcolor=blue, linkcolor=blue, citecolor=red}
\usepackage{xcolor}
\definecolor{newcolor}{rgb}{.8,.349,.1}

\journal{Journal of Computational Physics}

\usepackage{graphicx}
\usepackage{epsfig}
\usepackage{epstopdf}
\usepackage{multirow}
\usepackage{subfig}

\usepackage{bm}

\usepackage{todonotes}

\usepackage{amssymb,amsmath,array}

\biboptions{numbers}

\usepackage[inline]{enumitem}
\usepackage{siunitx}
\usepackage{algorithm}
\usepackage{algorithmicx}
\usepackage{algpseudocode}
\algrenewcommand\algorithmiccomment[1]{!!~{\itshape #1}}

\usetikzlibrary{arrows}

\begin{document}

\begin{frontmatter}

\title{Goal-based sensitivity maps using time windows and ensemble perturbations}%

\author[IC,NORMS]{C. E. Heaney} 
\corref{cor1}\ead{c.heaney@imperial.ac.uk}
\author[IC,NORMS]{P. Salinas} 
\author[IC,NORMS]{F. Fang}
\author[IC,NORMS]{C. C. Pain} 
\author[FL]{I. M. Navon} 

\cortext[cor1]{Corresponding author.}
\address[IC]{Applied Modelling and Computation Group, Department of Earth Science and Engineering, Imperial College 
London, UK}
\address[NORMS]{Novel Reservoir Modelling and Simulation Group, Department of Earth Science and Engineering, Imperial 
College London, UK}
\address[FL]{Department of Scientific Computing, Florida State University, Tallahassee, FL, 32306-4120, USA}

\begin{abstract}
We present an approach for forming sensitivity maps (or sensitivites) using ensembles. The method is an alternative 
to using an adjoint, which can be very challenging to formulate and also computationally expensive to solve. 
The main novelties of the presented approach are: 
1) the use of goals, by weighting the perturbation to help resolve the most important sensitivities, 
2) the use of time windows, which enable the perturbations to be optimised independently for each window and 
3) re-orthogonalisation of the solution through time, which helps optimise each perturbation when calculating 
sensitivity maps.  
The novel application of these methods, to the generation of sensitivity maps through ensembles, greatly reduces the 
number of ensemble members required to form the sensitivity maps as demonstrated in this paper.

As the presented method relies solely on ensemble members obtained from the forward model, it can therefore be 
applied directly to forward models of arbitrary complexity arising from, for example, multi-physics coupling, legacy 
codes or model chains. 

We demonstrate the efficiency of the approach by applying the method to advection problems and also 
a non-linear heterogeneous multi-phase porous media problem, showing, in all cases, that the number of 
ensemble members required to obtain accurate sensitivity maps is relatively low, in the order of 10s.

\end{abstract}


\end{frontmatter}

\section{Introduction}
\noindent 
Sensitivity maps are a key component of many numerical techniques found throughout computational physics, including 
mesh adaptivity~\cite{Power_2006}; correction of errors based on discretisation and/or modelling 
errors~\cite{Merton_2013, Merton_2014}; data assimilation based on adjusting the most sensitive parameters to the 
model-observation misfit~\cite{Cacuci_2005,Maday_2017,Shah_2017, Liu_and_Kalnay, Hossen_2012a}; and model optimisation~\cite{Cacuci_2012}. 
Sensitivity maps are also used to help identify the best locations and variables to measure for adaptive observations 
\cite{Dimet_2002, Daescu_2004, Che_2014}. Uncertainty in the physical measurement of parameters has meant that how a 
solution depends on small changes to model parameters (i.e.~sensitivities) is often as important as the solution 
itself~\cite{Cacuci_2013}. One example of this can be found in meteorological forecasting, where, to be 
useful, a forecast should be accompanied by an estimation of its likelihood~\cite{Leutbecher_2008}.

Sensitivities can be calculated by using first-order sensitivity analysis methods such as adjoints or direct 
sensitivity analysis (from perturbations)~\cite{Ionescu_2004}. Higher order sensitivity methods can be 
derived by the repeated application of first-order methods, as can be done when forming the Hessian 
matrix~\cite{Wang_1992}, for example. \citet{Cacuci_2015} has shown that a second-order adjoint can be formed with only 
a small amount of additional computational effort beyond that 
required to generate the forward model solutions and first-order sensitivities. However, adjoint models can be 
difficult to formulate for complex multi-physics problems and for new generation models that use adaptive mesh 
resolution. The Independent Set Perturbation adjoint method~\cite{Fang_2011} attempts to address these issues by 
forming an adjoint which can be incorporated into the computer code in a modular way, thereby aiding the 
maintainability of the code. The method can be applied to complex models on unstructured meshes, however, despite its 
modularity, introducing the code for the adjoint still requires substantial changes to be made to the 
original code. For multi-physics coupling, legacy codes or model chains, it is desirable to have a method 
\emph{completely} independent of the forward model. In which case, ensemble-based methods, such as the approach 
described here, offer a rather 
straightforward alternative to adjoint-based methods~\cite{Ancell_2007}. See, for instance, the approach of 
\citet{Jain_2018} based on Ensemble Kalman Filters (EnKF). Although conceptually more simple than 
adjoint-based models, ensemble-based models rely on solving the forward model multiple times. For complex models of 
realistic physics, it is therefore paramount to make the method as efficient as possible in order that the 
computational cost is not prohibitive. In this paper, we choose to minimise the ensemble size required to construct a 
sensitivity map of a `reasonable' accuracy, although other approaches exist. \citet{Li_2009} effectively reduce the 
computational cost of the forward model simulations by using Polynomial Chaos expansions, which allow the forward model 
simulations to be replaced by polynomial evaluations. Due to this reduction of computational effort, the ensemble size 
can be increased, which improves the accuracy of the method. In turbulent flow applications, the diagonally 
dominant state matrices mean that the covariance matrices in the data assimilation process can be approximated by banded 
matrices, thereby reducing the computational effort required~\cite{Meldi_2017}.

In this paper, we address how first-order sensitivity maps of high fidelity can be obtained with a minimal number of 
ensemble members, thereby increasing the computational efficiency of our approach. This high fidelity comes from a 
novel, integrated method including a new goal-based approach, in which the most up-to-date sensitivity maps are fed back 
into the perturbations to focus the algorithm on the key variables and domain areas, alongside smoothing and 
orthogonalisation of perturbations. Another key development is to introduce time windows (for time dependent problems) 
through which we work backwards in time in order to obtain greater accuracy in the sensitivity maps. Finally we report 
on how the solution can be re-orthogonalised in time to obtain sensitivity maps when the standard approaches fail 
(e.g.~when a very large number of ensemble members is used). 

In the goal-based approach presented here, the sensitivity of the functional with respect to the control variables is 
used to decide where to perturb. Thus, the perturbations that most affect the functional are selected. Similar, in 
some ways, to our approach, \citet{Keller_2010} uses the Leading 
Lyapunov Vector~\cite{Wolfe_2007} (LLV) that has the largest effect on the solution to optimise the initial 
perturbation. 
LLVs have the property that all random perturbations assume the characteristic structure of LLVs after a transient period. When several independent breeding cycles are performed, the phases and amplitudes of individual (and regional) LLVs are random, which ensures quasi-orthogonality among the global breeding vectors~\cite{Keller_2010, Norwood_2015}. 
Bred vectors (BVs) are aligned with the locally fastest growing LVs.
To generate BVs, at every breeding cycle, a
perturbation direction is determined and scaled by a small number. Next, it is added to the initial condition
of the control, in a similar manner to that described here, and the forward model is run. The result is subtracted 
from the unperturbed forward model results and the resulting perturbation is then used in the next BV, and so on, until 
the process converges. The BVs generated in this way will converge to the leading Lyapunov vector. 
BVs have been used to form sensitivities based on perturbations~\cite{Ancell_2007,Zupanski_2005,Zupanski_2006}, 
however, no one to date has used goal-based approaches to form 
the sensitivities. Other methods have used breeding methods (based on the LLVs or BVs methods)
\cite{Toth_1993,Toth_1997,Szunyogh_1997}, in which the perturbations excite the most energetic modes.
Alternatively, in weather forecasting, perturbations are often introduced guided by sensitivities from an 
adjoint~\cite{Cacuci_2005}. This is highly effective at targeting perturbations that grow quickly and can focus these 
on a particular goal. These sensitivities can be formed from an approximate sensitivity or from coarse grid/mesh models 
or reduced order models~\cite{Leroux_2018}. The same approach could be used here to form sensitivities quickly based 
on coarse models to reduce the overall computational cost. 

Optimising the accuracy of a goal to generate sensitivity maps is the key to  highly accurate performance, as has been 
discussed for goal-based error norms and mesh adaptivity~\cite{Pierce_2000,Power_2006, Merton_2013, Merton_2014}, 
and for sensor placement~\cite{Daescu_2004, Che_2014}.  
The optimisation provides the formal mathematical framework which is central to the 
success of these methods. Whereas such methods optimise the accuracy of a goal, we choose perturbations to form 
sensitivities such that the accuracy of the \emph{sensitivities} of the goal is optimised.  

Orthogonalisation is used here as it is in other perturbations methods, for EnKF~\cite{Attia_2016,Leroux_2018} 
and for optimal sensor placement~\cite{Keller_2010,Che_2014}. Orthogonalisation guarantees that the perturbations 
are independent, which enables the solution of the resulting systems of equations for the sensitivites and avoids 
duplication of effort.
 
As in other ensemble methods, smoothing is used to help form perturbations that avoid exciting the smallest 
scales and that tend to pick out the largest features quickly~\cite{Blum_2009, Nerger_2014, Attia_2016}. To achieve a 
similar result to smoothing, others have reduced the space of controls or parameters used in performing the sensitivity 
analysis to make running the forward model more manageable~\cite{Sun_2017}. 

Common to EnKF-based data assimilation~\cite{Attia_2016}, time windows are used here to focus the 
perturbations in such a way that the accuracy of the sensitivities to the perturbations at the start of each time 
window is maximised. In this way, the accuracy of the sensitivities in a time window can be extremely good with 
relatively few perturbations. Once the sensitivities of the final time window are calculated, this information can be 
passed back to the previous time window, recursively, until reaching the first time window. This approach is similar to 
adjoint methods which back-propagate the information from the goal to discover which variables affect the goal (and to 
what degree). The time window approach introduces the potential for parallelisation (albeit with some loss of 
speed of convergence) as all the time windows can be performed concurrently, unlike adjoint methods. Moreover, each 
perturbation within a time window can also be solved concurrently (again with some loss of convergence speed), making 
the combination very appealing for parallelisation.
This can make the generation of sensitivity maps extremely efficient for large problems. Nonetheless, the 
parallelisation is not explored in this paper and is left as a future area of research. 

The goal-based framework, the use of time windows and orthogonalisation provide the focus for this paper; the remainder 
of which is as follows. Section~\ref{sec:ensemble_theory_background} outlines the ensemble theory. 
Section~\ref{sec:ensemble_theory_new} describes the goal-based approach of using the sensitivities to focus the 
perturbations; the use of 
time windows; and the re-orthogonalisation of the solution through time. Section~\ref{Results} develops~1D and~2D 
advection examples and a 3D multi-phase flow problem. These example problems are chosen to have incrementally increased 
complexity in order to explore the performance of the developed methods. In the final section conclusions are drawn.

\section{Ensemble theory - background}\label{sec:ensemble_theory_background}
Central to the following theory is the calculation of the sensitivity of a functional with respect to the solution 
variables. We therefore begin this section by showing how the sensitivities influence the change in 
the value of a functional~\cite{Merton_2013}. We then explain how ensembles can be used to approximate sensitivies, 
and extend this to time-dependent problems before describing several techniques found in the literature for generating 
ensemble members.

Suppose we have a set of control parameters $\bm{m}$ that are inputs to a computational model and that are related to a 
set of unperturbed control parameters $\overline{\bm{m}}$ by the relationship 
\begin{equation}
\bm{m} =  \overline{\bm{m}} + \Delta \bm{m} \,,
\end{equation}
where $\Delta \bm{m}$ represents a small perturbation. Although, in this paper, initial conditions are used 
for the controls, other quantites could be used such as boundary conditions, sources or model parameters. Associated 
with the perturbed controls are solution variables~$\bm{\Psi}$ and a smooth (differentiable) 
functional~$F(\bm{\Psi}(\bm{m}),\bm{m})$. The 
solution variables and functional value differ from their unperturbed counterparts ($\overline{\bm{\Psi}}$ and 
$\overline{F}$) as 
follows:
\begin{eqnarray}
\Delta \bm{\Psi} & = & \bm{\Psi} - \bm{\overline{\Psi}} \\[1mm]
\Delta F & = & F - \overline{F}\,. 
\end{eqnarray}
We approximate the effect that a small change in the controls has on both the solution variables and the functional by 
first order Taylor series expansions 
\begin{eqnarray}
\Delta \bm{\Psi} & = &  \frac{{\rm d} \bm{\Psi}}{{\rm d} \bm{m}} \Delta \bm{m} =: \bm{M} \Delta \bm{m} 
\label{delta-psi} \\[2mm]
\Delta F & = & {\Delta \bm{m}}^T \frac{\partial F}{\partial \bm{m}} + {\Delta \bm{\Psi}}^T \frac{\partial F}{\partial 
\bm{\Psi}} \  = \ {\Delta \bm{m}}^T \frac{{\rm d} F}{{\rm d} \bm{m}} \,. \label{delta-F}
\end{eqnarray}
The matrix $\bm{M}$, defined in equation~\eqref{delta-psi}, has values
\begin{equation}
\bm{M}_{ij}=\frac{{\rm d} \bm{\Psi}_i}{{\rm d} {\bm{m}}_j}, \; \forall i\in\{1,2,\ldots,{\cal N}\}\,, \; 
\forall j\in\{1,2,\ldots,{\cal C}\}\,, 
\end{equation}
where $\cal N$ is the total number of solution variables (e.g.~the number of control volumes 
or finite element nodes) and $\cal C$ is the number of controls. To relate a change in the solution variables to 
a change in the functional, first, we rearrange equation~\eqref{delta-psi} to obtain an expression for $\Delta\bm{m}$ 
in terms of $\Delta\bm{\Psi}$ and, second, substitute this into equation~\eqref{delta-F}. Rearranging for 
$\Delta\bm{m}$ requires the use of the Moore-Penrose pseudo-inverse, as $\bm{M}$ is, in general, a non-square matrix:
\begin{equation}\label{eq:delta_m}
\Delta \bm{m} = (\bm{M}^T \bm{M})^{-1} \bm{M}^T \Delta \bm{\Psi}\,.
\end{equation}The matrix  $(\bm{M}^T \bm{M})$ will have full rank provided the vectors that make up the columns of 
the matrix ${\bm{M}}$ are linearly independent. Substituting equation~\eqref{eq:delta_m} into 
equation~\eqref{delta-F} yields 
\begin{equation}\label{delta-F2}
\Delta F= {\Delta\bm{\Psi}}^T \left( \bm{M} (\bm{M}^T \bm{M})^{-1} \frac{{\rm d} F}{{\rm d} \bm{m}} \right) =: 
{\Delta\bm{\Psi}}^T 
\bm{g}\,,
\end{equation}
where we define $\bm{g}$ to be the quantity which relates a change in the solution variables to a change in the 
associated functional. Once $\bm{g}$ is known, given an estimate for ${\Delta\bm{\Psi}}$, the influence of each 
solution variable on $\Delta F$ can be determined. One method for estimating ${\Delta\bm{\Psi}}$ is to use 
interpolation theory~\cite{Pain_2001}. By further inspection of equation~\eqref{delta-F2}, we see that the term in 
parentheses, defined as $\bm{g}$, is, in fact, the total derivative of the functional with respect to the solution 
variables
\begin{equation}
\bm{g} \equiv \frac{{\rm d} F}{{\rm d} \bm{\Psi}}\,,  
\label{d-def-adjoint}
\end{equation}
thus $\bm{g}$ is equivalent to the sensitivity of the functional with respect to the solution variables.
Henceforth, we refer to $\bm{g}$ as a sensitivity map.

\subsection{The use of ensembles to form approximate sensitivities}\label{subsec:ensembles4sensitivities}
Suppose we now have a total of $\cal{E}$ ensemble members, each with a set of perturbed controls denoted by 
${}^e\bm{m}$ which differs from the unperturbed controls by $\Delta{}^e\bm{m}$, where $e$ corresponds to the particular 
ensemble.
We wish to perform a sensitivity analysis around the variables $\overline{\bm{\Psi}}$ and $\overline 
F$ associated with an unperturbed set of controls $\overline{\bm{m}}$. Before doing so, it is convenient to make a 
change of variables which maps the controls to a space with a reduced number of variables. Therefore, we define a 
second 
vector of control variables $\Delta{}^e\bm{m}_s$ of length equal to the number of ensemble members (or 
ensemble size), $\cal{E}$, whose entries are either zero or one:
\begin{equation}\label{eq:change_of_variables}
(\Delta{}^e\bm{m}_{s})_{k} = \left\{ \begin{array}{l} 0 \quad \text{for } k\in\{1,\ldots,{\cal E}\}, k\neq e \\ 1 
\quad \text{for } k = e\,.\end{array} \right.
\end{equation}
The perturbations in both variables are related through 
\begin{equation}
\bm{C} \Delta {}^e\bm{m}_s = \Delta {}^e\bm{m} \quad\forall\,e\,,
\label{Cm} 
\end{equation}
where $\bm{C}_{ie}=\, \Delta{}^e\bm{m}_i$.  

Repeating the sensitivity analysis described in equations~\eqref{delta-psi}, \eqref{delta-F}, \eqref{eq:delta_m} 
and~\eqref{delta-F2} in the new variables, i.e.~using $\bm{m}_s$ rather than $\bm{m}$, yields
\begin{equation}
\Delta F =  \Delta \bm{m}_s^T \frac{{\rm d} F}{{\rm d} \bm{m}_s} = {\Delta \bm{\Psi}}^T \left( \bm{M}_s  
(\bm{M}_s^T \bm{M}_s)^{-1}  \frac{{\rm d} F}{{\rm d} \bm{m}_s} \right) = {\Delta \bm{\Psi}}^T \bm{g}\,,
\label{key-mod-eqn} 
\end{equation}
in which $\bm{M}_s$ is now of size~$\cal{N}$ by~$\cal{E}$ where, most likely, $\cal{E}\ll\cal{C}$. 
We remark that, by applying the chain rule to the derivative of the functional with respect to the controls, 
\begin{equation}
\frac{{\rm d} F}{{\rm d} \bm{m}_s} = \left( \frac{{\rm d} \bm{\Psi}}{{\rm d} \bm{m}_s}\right)^T 
\frac{\partial F}{\partial \bm{\Psi}} + \frac{\partial F}{\partial \bm{m}_s}\,, 
\label{dF-dms}
\end{equation}
reveals that ${\rm d} \bm{\Psi} /{\rm d} \bm{m}_s$ is the approximate adjoint operator. 
In order to calculate the sensitivity map in the new variables, we need to approximate $\bm{M}_s$ and 
${\rm d} F / {\rm d} \bm{m}_s$ (see equation~\eqref{key-mod-eqn}). Since the perturbations have been defined to have 
a magnitude of one in the new variables we can make the following approximations:
\begin{eqnarray}
\frac{{\rm d} F}{{\rm d} \bm{m}_s} & \approx & \widehat{\frac{{\rm d} F}{{\rm d} \bm{m}_s}} = ( {}^{1}\!{F} - 
\overline F, {}^2\!F - \overline F, \ldots,  {}^{\cal E}\!F - \overline F )^T
\label{hat-F} \label{eqn:approxF} \\[2mm]
\bm{M}_s = \frac{{\rm d} \bm{\Psi}}{{\rm d} \bm{m}_s}  & \approx & {\widehat{\bm{M}}_s} = 
(^1\bm{\Psi}-\overline{\bm{\Psi}},  \; ^2\bm{\Psi}-\overline{\bm{\Psi}}, \; 
\ldots\; ,\;^{\cal E}\bm{\Psi} -\overline{\bm{\Psi}} ). 
\label{widehat M}
\end{eqnarray}
Substituting these approximations into equation~\eqref{key-mod-eqn} yields
\begin{equation}
\widehat{\Delta F} = \widehat{\Delta \bm{\Psi}}^T \left( {\widehat{\bm{M}}_s} ({\widehat{\bm{M}}_s}^T 
{\widehat{\bm{M}}_s})^{-1}  \widehat{\frac{{\rm d} F}{{\rm d} \bm{m}_s}} \right) = \widehat{\Delta \bm{\Psi}}^T 
\widehat{\bm{g}}\,,
\label{key-mod-eqn-final1} 
\end{equation}
where $\widehat{\Delta F} \approx \Delta F$ and 
$\widehat{\bm{g}}\approx \bm{g}$. The approximation of the sensitivity map can be written explicitly as
\begin{equation}
\widehat{\bm{g}}={\widehat{\bm{M}}_s} \left({\widehat{\bm{M}}_s}^T {\widehat{\bm{M}}_s}\right)^{-1} 
\widehat{\frac{{\rm d} F}{{\rm d} \bm{m}_s}}\,. 
\label{key-mod-eqn-final-g} 
\end{equation}
The benefit of the change of variables is that neither $\widehat{{\rm d} F / {\rm d} \bm{m}_s}$ nor 
${\widehat{\bm{M}}_s}$ depend on the controls, $\bm{m}_s$, as 
$\bm{\Psi}$ and $F$ are invariant under the change of 
variables, see equations~\eqref{eqn:approxF} and~\eqref{widehat M}. Consequently the matrix $\bm{C}$ does not appear in 
the expression for the sensitivity map and never 
needs to be constructed. Notice that ${\widehat{\bm{M}}_s}^T {\widehat{\bm{M}}_s}$ is a square matrix of order equal to 
the ensemble size~$\cal E$ which may make tractable its inversion, required for the calculation of the 
sensitivity map. 

Further examination of equation~\eqref{key-mod-eqn-final1} shows that each contribution to~$\widehat{\Delta F}$ is of 
the form
\begin{equation}\label{delta-F-final} 
 \widehat{\Delta\bm{\Psi}}_i\,\widehat{\bm{g}}_i\,.
\end{equation}
Thus, the magnitude of the $i^{th}$ entry of the sensitivity map determines how significant the influence is of the 
$i^{th}$ solution variable upon the functional. Little information will be gained, therefore, by perturbing the 
controls at nodes (or in control volumes) where the value of $\widehat{\bm{g}}$ is close to zero, whereas more will be  
gained by perturbing the controls where the solution variables have a greater effect on the functional, i.e.~where 
there is a high value of~$\widehat{\bm{g}}$. Therefore equation~\eqref{delta-F-final} provides the motivation to bias 
the perturbations with the sensitivity map in the goal-based approach outlined later, in 
section~\ref{sec:orthog_ensemble_members_biasing}. By adopting this approach we are 
effectively choosing the perturbations that most affect the functional~$F$.

\subsection{Time dependent problems}
\par\noindent
Time dependent problems can also be treated with this method by directly applying the approach outlined in 
section~\ref{subsec:ensembles4sensitivities} simultaneously in space and time. To do this, we would construct 
$\widehat{\bm{\Psi}}$ so that it contained all the solutions in space and time, $\widehat{\bm{M}}_s$ so that it 
contained all derivatives in space and time etc. However, the accuracy can be enhanced by considering each time level 
separately. 
In order to do this, we apply to equation~\eqref{delta-psi} the change of variables given in 
equation~\eqref{eq:change_of_variables} and the approximations given in equations~\eqref{eqn:approxF}, \eqref{widehat M} 
and following~\eqref{key-mod-eqn-final1}, resulting in
\begin{equation}\label{eq:4modified}
 \widehat{\Delta \bm{\Psi}} = \widehat{\bm{M}}_s \Delta\bm{m}_s\,.
\end{equation}
We then construct $\widehat{\bm{M}}_s$ and $\widehat{\bm{\Psi}}$ as below having substituted these expressions into 
equation~\eqref{eq:4modified} to give
\begin{equation}\label{eq:space-and-time}
\left( (\widehat{\Delta\bm{\Psi}}^{1})^T, \ldots,(\widehat{\Delta\bm{\Psi}}^{{\cal N}_t})^T \right)^T  = \left( 
(\widehat{\bm{M}}_s^{1})^T,\ldots,(\widehat{\bm{M}}_s^{{\cal N}_t} )^T \,\right)^T   
\Delta \bm{m}_s
\end{equation}
in which the time level is indicated as a superscript and ${\cal N}_t$ is the number of time levels over the  
interval $t\in\left[0,\Delta t{\cal N}_t\right]$. We remark that the mapping represented by the matrix~$\bm{C}$ in 
equation~\eqref{Cm} is different for each time level. The system of equations~\eqref{eq:space-and-time} can now be 
decoupled in time and, once the Moore-Penrose pseudo-inverse is applied to the equation for each time level in turn, 
we have
\begin{equation}\label{eq:g_n}
\widehat{\bm{g}}^{n} = \widehat{\bm{M}}^{n}_s \left( (\widehat{\bm{M}}^{n}_s)^T \widehat{\bm{M}}^{n}_s \right)^{-1} 
\widehat{ \frac{{\rm d} F}{{\rm d} \bm{m}^n_s}  }\,,
\end{equation}
for time level $n$. The interpretation of $\widehat{ \frac{{\rm d} F}{{\rm d} \bm{m}^n_s}  }$ requires some 
explanation. Assuming that the functional depends on the controls only through the solution we can write
\begin{equation}
\widehat{F} = \widehat{F}(\bm{\Psi}^1, \bm{\Psi}^2, \ldots, \bm{\Psi}^{\mathcal{N}_t})\,.
\end{equation}
If a perturbation is made at time~$t^n$, only solutions after this time will be affected. When evaluating the 
functional in these cases, the solutions before time~$t^n$ will be identical to the unperturbed solutions 
the unperturbed solutions at times before~$t^n$
\begin{equation}
\widehat{F} |^n = \widehat{F}(\overline{\bm{\Psi}}^1, \overline{\bm{\Psi}}^2, \ldots, \overline{\bm{\Psi}}^{n-1}, 
\bm{\Psi}^n, \bm{\Psi}^{n+1}, \ldots, \bm{\Psi}^{\mathcal{N}_t})\,.
\end{equation}
The quantity $\overline{F}$ can be written as
\begin{equation}
\overline{F} = \overline{F}(\overline{\bm{\Psi}}^1, \overline{\bm{\Psi}}^2, \ldots, 
\overline{\bm{\Psi}}^{\mathcal{N}_t})\,.
\end{equation}
For ensemble member $e$, applying the same reasoning as before (see equation~\eqref{eqn:approxF}), the derivative of the 
functional with respect to a perturbation in the controls at time level~$n$ can be approximated by
\begin{equation}
\left({}^e\widehat{F} - \overline{F} \right) \Big|^n\,.
\end{equation}
Suppose, for example, that the functional were the integral in time of the solution at node~$j$ approximated as follows
\begin{equation}
 {}^e\widehat{F} = \sum_{k=0}^{\mathcal{N}_t} \Delta t^k \  {}^e\bm{\Psi}^k_j\,,
\end{equation}
then, if the perturbations are made at time level~$n$, we have 
\begin{equation}
\left({}^e\widehat{F} - \overline{F} \right) \Big|^n =  \sum_{k=n}^{\mathcal{N}_t} \Delta t^k \, \left(  
{}^e\bm{\Psi}^k_j - \overline{\bm{\Psi}}^k_j \right)\,,
\end{equation}
that is, no contribution is made from terms relating to times earlier than~$t^n$. We can now extend 
equation~\eqref{eqn:approxF} for time dependent problems and approximate the derivative of the functional with 
respect to the controls as follows
\begin{equation}
 \widehat{ \frac{{\rm d} F}{{\rm d} \bm{m}^n_s}  } = \left( \left({}^1\widehat{F} - \overline{F} \right) \Big|^n, 
\left({}^2\widehat{F} - \overline{F} \right) \Big|^n, \, \ldots\,,\, \left({}^{\mathcal{E}}\widehat{F} - \overline{F} 
\right) \Big|^n \right)^T\,.
\end{equation}

\subsection{Methods for the generation of ensembles}
\label{Method of generating the ensembles}
\par\noindent
We describe here three common techniques that can be applied when perturbing fields which will later be used to 
generate ensemble members: random perturbations, smoothing and orthogonalisation. 

\subsubsection{Random perturbations}\label{sec:random_pert}
The advantage that the use of random perturbations has over independently perturbing each degree of freedom in turn, is 
that random perturbations will excite all solution modes simultaneously. This can be much more efficient, requiring 
fewer ensemble members in order to obtain a converged sensitivity map. Also, almost always (for truly random numbers), 
using 
random fields will produce a linearly independent set of perturbations which means that the matrix 
$\widehat{\bm{M}}_s^T \widehat{\bm{M}}_s$, at the initial time, has full rank and can be inverted.

\subsubsection{Smoothing}\label{sec:smoothing}
Smoothing is applied in order to remove small-scale grid noise without affecting the underlying physical 
structure~\cite{Shapiro_1970}. The smoothed perturbations are able to pick out large scale structures in the 
sensitivities faster than non-smoothed random perturbations. The reason for this is that the perturbations 
cover relatively large regions of the domain (compared to the grid/mesh spacing). Furthermore, the smoothing is key to 
reducing the number of ensemble members required. We smooth the resulting vector of perturbations $\Delta\bm{m}$ as 
follows:  
\begin{equation}\label{eq:smoothing_S}
\Delta \bm{m} \rightarrow {\bm{S}} \Delta \bm{m}\,. 
\end{equation}
The matrix ${\bm{S}}$ defines the smoothing and is given by 
\begin{equation}
\bm{S}_{ii} =\frac{1}{2}\,,\ 
\bm{S}_{ij} = \left\{ \begin{array}{cl} \dfrac{1}{2 v_i} & \text{for each neighbouring cell~$j$ of cell~$i$}\\[3mm]
                                            0  & \text{for each non-neighbouring cell~$j$ of cell~$i$} 
\end{array}\right. 
\end{equation}
 where the valency~$v_i$ of cell~$i$ is the number of cells next to cell~$i$. (Read node(s) for cell(s) if the finite 
element method is used.) The number of smoothing steps is chosen here as the nearest integer value to a quarter of the 
maximum number of cells across the domain in any direction.  

\subsubsection{Orthogonalising with respect to the previous ensemble members}\label{sec:orthog_and_scale}
After generating a vector of random perturbations for the $e^{th}$ ensemble, $\Delta {}^{e}\bm{m}$, we  
orthogonalise with respect to the previous members by using the standard Gram-Schmidt orthogonalisation process below:
\begin{equation}
\Delta {}^{e} \bm{m} \rightarrow \Delta {}^{e} \bm{m} - \sum_{k=1}^{e-1} \frac{
(\Delta {}^{k} \bm{m})^T \Delta {}^{e} \bm{m} }{(\Delta {}^{k} \bm{m})^T \Delta {}^{k} \bm{m}}\,\Delta {}^{k} 
\bm{m}\,.
\label{othog}
\end{equation}

Finally, we rescale the modified perturbation by $L$ and a small number, $\epsilon$, (e.g.~$\epsilon 
=10^{-4}$): 
\begin{equation}
\Delta {}^{e}\bm{m} \rightarrow  \frac{\epsilon}{L}\,\Delta {}^{e}\bm{m} \,, 
\label{eq:rescale}
\end{equation}
\noindent so that every perturbation is given equal priority. For the simple 1D cases we set~$L$ to be such that the 
orthogonalisation step~\eqref{othog} and the scaling step~\eqref{eq:rescale} are equivalent to Gram-Schmidt 
orthonormalisation. For the more complex~2D and 3D cases we set $L$ as given below
\begin{equation}
L=\left\{ \begin{array}{ll} \sqrt{(\Delta {}^{e}\bm{m})^T \Delta {}^{e}\bm{m}}  & \text{for 1D} \\[2mm] \max\{\Delta 
{}^{e}\bm{m}\} - \min\{\Delta {}^{e}\bm{m}\}& \text{for 2D and 3D.} \end{array} \right.
\end{equation}

\section{Ensemble theory - new developments}\label{sec:ensemble_theory_new}
\noindent

In this section we introduce three novel contributions: weighting the perturbations by the sensitivity map, the use 
of time windows and re-orthogonalising through time. Common to the second and third contributions is that, at certain 
points in time, new perturbations are calculated and the simulation is resumed. 

\subsection{Goal-based weighting of the perturbations}\label{sec:orthog_ensemble_members_biasing}
One may weight the perturbations based on the current value of $\widehat{\bm{g}}^k$ at time level zero, (i.e.~$k=0$) to 
focus the perturbations in the regions more relevant to the goal of interest: 
\begin{equation} 
\Delta \bm{m} \rightarrow 
\frac{\vert\widehat{\bm{g}}^k\vert \odot \Delta \bm{m}}{\vert\vert\widehat{\bm{g}}^k\vert\vert_\infty 
}\,,
\label{g-bias} 
\end{equation}
where $\odot$ represents the Hadamard product (element-wise multiplication) and $\vert\vert\cdot\vert\vert_{\infty}$ 
represents the infinity or maximum norm. By current, we mean that $\widehat{\bm{g}}^k$ is calculated using 
information from as many ensemble members as have been created by this point. In this way, the perturbations will be 
focused 
on the areas in the domains and the variables that have most impact on the functional. This approach is therefore 
analogous to goal-based mesh adaptivity~\cite{Power_2006}, and goal-based sensor optimisation methods~\cite{Che_2014}, 
but instead of refining the mesh or observations within the sensitive areas and variables, it refines the perturbations 
within these.  
We include weighting in our algorithm for generating the perturbations along with methods already in the literature 
as 
follows:
\begin{itemize}
   \item create uniformly distributed random perturbations (see section~\ref{sec:random_pert})
   \item apply smoothing (see section~\ref{sec:smoothing})
   \item weight the perturbations with a sensitivity map based on however many ensemble members are available at this 
point
   \item apply Gram-Schmidt orthogonalisation with respect to previous ensemble members and scale 
(section~\ref{sec:orthog_and_scale})        
\end{itemize}
At this point the modified perturbations are added to the unperturbed initial conditions, following which, the forward 
model is solved yielding the solution variables and functional. The smoothing, weighting, and orthogonalising and 
scaling steps all describe modifications to the random perturbations which are designed to increase the efficiency of 
the method and reduce the ensemble size required. 

\begin{figure}[h!]
  \centering
    \includegraphics[width=0.9\textwidth]{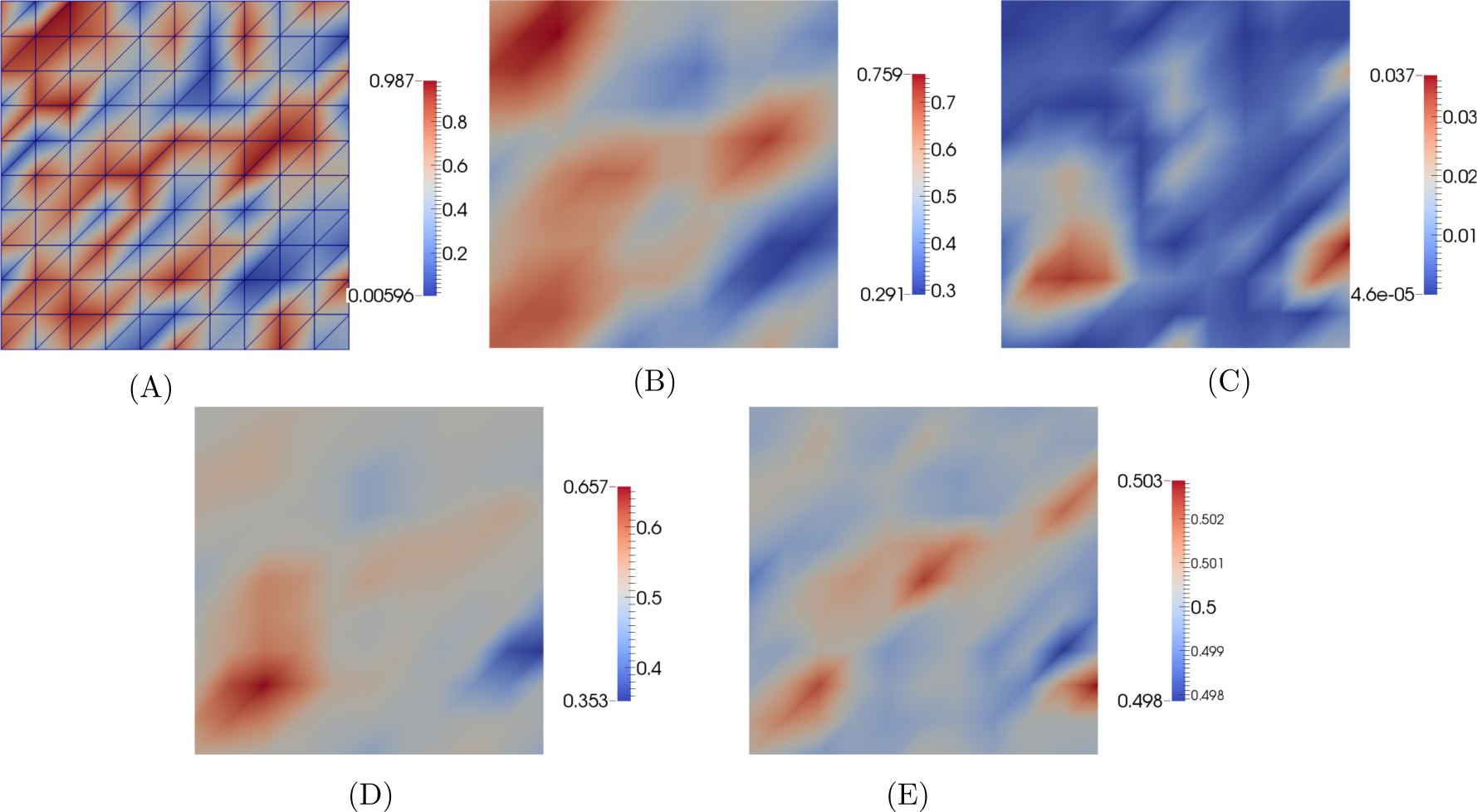}
    \caption{(A) The initial uniformly distributed random perturbation. (B) After applying smoothing to (A). (C) 
Sensitivity map based on 10 ensemble members. (D) Initial perturbation after smoothing and weighting by the importance 
map (C). (E) Initial perturbation after smoothing, weighting and orthogonalisation with respect to previous 
ensemble members and 
re-scaling. 
\label{fig:perturbation_steps}}
\end{figure}

These steps are illustrated in Figure~\ref{fig:perturbation_steps}, where we see  
(A) uniformly distributed perturbations added to 
 the unperturbed initial conditions, (B) smoothing applied to 
this, (C) the importance map (after 10~ensemble members) at the 
 initial time, (D) the perturbed initial condition weighted 
by the importance map and (E) the initial condition having 
 been orthogonalised with respect to previous ensemble members 
and re-scaled. 

\subsection{Use of time windows, working backwards through time and 
concurrency}\label{sec:time_windows} 
We split up the time domain into a number of non-overlapping intervals referred to as time windows, 
ensuring that there is an integer number of time steps within each time window. Once the unperturbed solution is 
available throughout time, the controls are also known at the beginning of 
each time window and are perturbed, thereby providing an `initial condition' for each window. 
In this way a separate set of perturbations is applied to each time window. Due to these different sets of 
perturbations, the derivative $\frac{ {\rm d} F}{ {\rm d} \bm{\Psi}}$ (i.e.~the sensitivity map) has to be mapped 
backwards in time through the time windows taking account of the change in the controls, $\bm{m}_s$ in each window. 
This mapping is now described.

\begin{figure}[htbp!]
 \centering
 \scalebox{0.8}{\includegraphics[trim=30mm 145mm 30mm 30mm, clip]{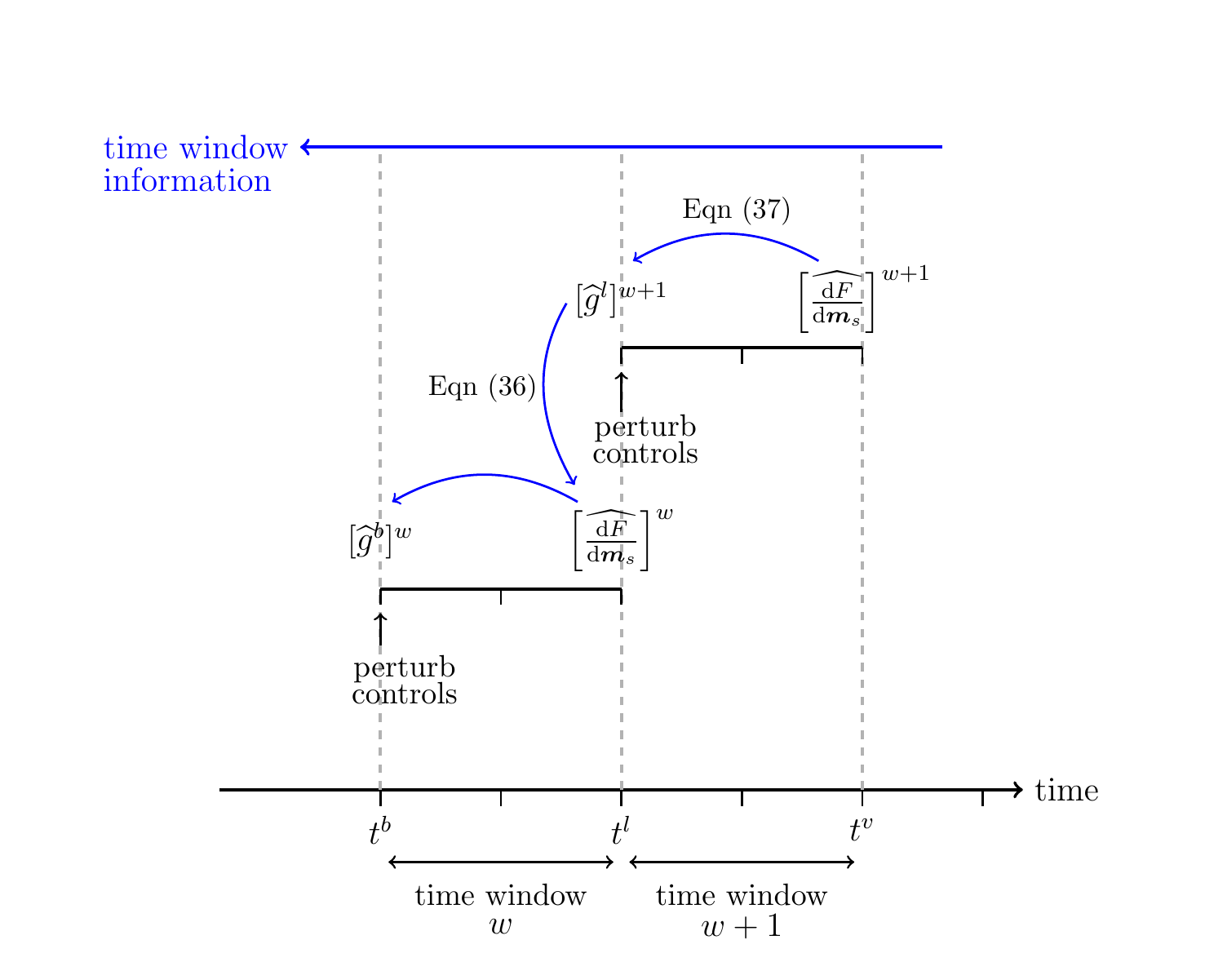}}
\caption{Two time windows are shown, window $w$, in which $t\in[t^b,t^l)$ and window $w+1$, in which $t\in[t^l,t^v)$ 
where $t^b$, $t^l$ and $t^v$ are global time levels. This illustrates the process of calculating the derivative 
$\frac{\widehat{\rm{d} F}}{\rm{d} \bm{m}^l_s}$ for window $w$ based on $\widehat{\bm{g}}^l$ of time window $w+1$, 
which is already known at this point due to the time windows being passed through in reverse order. }
\label{fig:time_window_diagram}
\end{figure}

Suppose that window $w$ is the time interval $[t^b, t^l)$, and window $w+1$ is the next time interval $[t^l, 
t^v)$,  where $b$, $l$ and $v$ are indices that denote (global) time levels. Using the chain rule we can evaluate the 
derivative $\frac{ {\rm d} F}{ {\rm d} {\bm{m}}_s^l}$ in the current window, $w$, as follows
\begin{equation}
\left[\frac{ {\rm d} F}{ {\rm d} {\bm{m}}_s^l}\right]^w=
 \left[\frac{ {\rm d} \bm{\Psi}^l}{ {\rm d} {\bm{m}}_s^l}\right]^w  
\left[\frac{ {\rm d} F}{ {\rm d} {\bm{\Psi}}^l}\right]^{w+1}\,.
\end{equation}
Applying the approximations we have used earlier (in equations~\eqref{eqn:approxF}, \eqref{widehat M} and 
following~\eqref{key-mod-eqn-final1}) at time level $l$ gives
\begin{equation}\label{ll}
\left[ \widehat{\frac{{\rm d} F}{ {\rm d} {\bm{m}}_s^l}} \right]^w=
\left[ (\widehat{ {\bm{M}} }^l_s)^T \right]^w \left[ \widehat{\bm{g}}^l\right]^{w+1} \,,
\end{equation}
where $\widehat{\bm{g}}^l$ has already been calculated in window~$w+1$. For time window~$w$, we can write the 
sensitivity map as follows

\begin{equation}\label{g_for_time_windows-total-deriv}
\widehat{\bm{g}}^{k} = {\widehat{\bm{M}}_s^{k}} \left( (\widehat{\bm{M}}_s^{k})^T \widehat{\bm{M}}_s^{k} 
\right)^{-1} 
\left(\left[ \widehat{\frac{{\rm d} F}{{
\rm d} {\bm{m}}_s^l}} \right]^w + 
 \frac{{\rm d} F}{{\rm{d}} {\bm{m}}_s^k} 
\right)  , \;\; \forall k\in\left[b,l\right). 
\end{equation}
or using:
\begin{equation}\label{total-derive}
\frac{{\rm{d}} F}{{\rm{d}}{\bm m}_s^k} = \sum_{q=k}^{l-1} \frac{{\rm{d}} {\bm\Psi}^q}{{\rm{d}} {\bm{m}}^k_s} \frac{\partial F}{\partial {\bm\Psi}^q}
+\frac{\partial F}{\partial {\bm{m}}^k_s },
\end{equation}
then: 
\begin{equation}\label{g_for_time_windows-}
\widehat{\bm{g}}^{k} = {\widehat{\bm{M}}_s^{k}} \left( (\widehat{\bm{M}}_s^{k})^T \widehat{\bm{M}}_s^{k} 
\right)^{-1} 
\left( 
\left[\widehat{ \frac{{\rm d} F}{{\rm d} {\bm{m}}_s^l}  }\right]^w
+ 
\sum_{q=k}^{l-1} (\widehat{ {\bm{M}} }^q_s)^T \frac{\partial F}{\partial {\bm{\Psi}}^q}
+\frac{\partial F}{\partial {\bm{m}}^k_s }\right), \;\; \forall k\in\left[b,l\right), 
\end{equation}

where all terms are now associated with time window~$w$, hence the index~$w$ is omitted. See 
Figure~\ref{fig:time_window_diagram} for an illustration of this process. We show that using time windows can improve 
immensely the fidelity of the sensitivity maps for a given number of perturbations. This is because the perturbations 
can be more strongly focused on resolving the sensitivities in a particular time window, which results in the need for 
substantially fewer perturbations, see section~\ref{Results}. 

Another key attribute of the time windows method is that each time window can be performed in isolation leading to 
``explicit time windows''. Using explicit time windows will have a certain loss of accuracy as 
$\widehat{\bm{g}}$ is no longer used to guide the generation of perturbed ensemble members. Nonetheless, explicit time 
windows 
introduce the potential to exploit parallelisation, as not only the ensemble members can be solved in 
parallel, but also the different time windows can be performed concurrently, enabling massive parallelisation.


For `short' time windows or when relying on a small number of perturbations, it may be useful to weight the 
perturbations with the previously calculated value of the sensitivity map from a future time window. This serves as an 
estimate of the initial value of $\widehat{\bm{g}}$ for the current time window as in the previously described 
goal-based approach (see section~\ref{sec:orthog_ensemble_members_biasing}). For example, in time window $w$, the 
sensitivity map at the initial time of window~$w$, $\widehat{\bm{g}}^b$, could be approximated by the sensitivity map 
at the initial time of window~$w+1$, i.e.
\begin{equation}
\left[ \widehat{\bm{g}}^b \right]^{w} \approx \left[ \widehat{\bm{g}}^l \right]^{w+1}\,.
\end{equation}

\subsection{Re-orthogonalising through time}\label{sec:re-orthogonalising-through-time}
It is advantageous to orthogonalise the ensemble members for several reasons: 
 (a) Each ensemble member will be given approximately equal priority in the 
generation of the sensitivity maps; 
 (b) An independent set of ensemble members is guaranteed. Some random number generators have 
biases in them which may produce non-independent ensemble sets. This results in poor conditioning 
of the matrices that we invert within the procedure;
 (c)  One can lose independence of the sets as time evolves, which means that 
some of the ensemble members are not contributing to the accuracy of the calculated $\widehat{\bm{g}}$. One way of addressing this issue 
is to use regularisation when forming 
the Moore-Penrose pseudo-inverse. That is introducing a positive diagonal 
matrix 
\begin{equation}
\left(\widehat{\bm{M}}_s^T\widehat{\bm{M}}_s \right)^{-1} 
\rightarrow 
\left(\widehat{\bm{M}}_s^T\widehat{\bm{M}}_s +\epsilon_s {\bf I}\right)^{-1}
\label{reg-I}
\end{equation} 
 for some small $\epsilon_s$.  An example of an equation that may be used to calculate 
$\epsilon_s$ is:
\begin{equation}
\epsilon_s =  \frac{\alpha_s}{\cal E} \sum_{k=1}^{\cal E} \left(\widehat{\bm{M}}_s^T\widehat{\bm{M}}_s \right)_{kk}
\end{equation} 
in which $\alpha_s$ is a small number. We have used $\alpha_s=1\times 10^{-14}$ 
in the 1D applications where we have not used 
re-orthogonalisation or time windows. 
An alternative to using regularisation (equation~\eqref{reg-I}) 
is to re-orthogonalise the sets as the simulation progresses in time, as explained here.

We consider re-orthogonalising every time step in order to avoid losing independence of the set of ensemble members 
and thus being unable to invert the matrix $\widehat{\bm{M}}_s^T\widehat{\bm{M}}_s$. In practice one may 
re-orthogonalise every~$N$ time steps (where~$N> 1$) to reduce the burden of this computation.

For a one level time discretisation, the solution at a particular time level can be thought of as the initial condition 
for the next time level, hence in this section the vector $\bm{m}$ containing the `initial conditions' will now have a 
time index and satisfies
\begin{equation}
  \bm{m}^n = \bm{\Psi}^n\,,
\end{equation}
where $\bm{m}^n$ are the initial conditions which would lead to the discrete solution~$\bm{\Psi}^{n+1}$.
Suppose we have a solution vector 
at time level~$n$, $\bm{\Psi}^n$. We set the deviation of this value from the unperturbed solution as the 
perturbation of the initial condition for the solution at the next time level
\begin{equation}
\Delta \bm{m}^n = \Delta\bm{\Psi}^n = \bm{\Psi}^n - \overline{\bm{\Psi}}^n\,.
\end{equation}
It is at this point that we orthogonalise with respect to all other ensemble values at this time level
\begin{equation}
\Delta {}^{e}\tilde{\bm{m}}^n  = \Delta {}^{e} \bm{m}^n - \sum_{k=1}^{e-1} \frac{
(\Delta {}^{k} \bm{m}^n)^T \Delta {}^{e} \bm{m}^n }{(\Delta {}^{k} \bm{m}^n)^T \Delta {}^{k} \bm{m}^n}\,\Delta {}^{k} 
\bm{m}^n\,.
\end{equation}
As before, for convenience, we wish to use the change of variables and so must also consider $\Delta 
{}^e\tilde{\bm{m}}^n_s$. Assuming, ${\Delta{}^e\bm{m}}_s^n$ is the set of perturbed `initial' conditions at time 
level~$n$ and $\Delta{}^e\tilde{\bm{m}}_s^n$ the orthogonalised set, then by definition
\begin{equation}
 \Delta {}^e\tilde{\bm{m}}^n_s =  \Delta {}^e\bm{m}^{n+1}_s\,,
\end{equation}
which leads to the result
\begin{equation}\label{eq:magic}
\widehat{\frac{{\rm d} F}{{\rm d} \tilde{\bm{m}}_s^n }} = \widehat{\frac{{\rm d} F}{{\rm d} {\bm{m}}_s^{n+1} }}\,. 
\end{equation}
This is true because, at each time step, the change of variables is such that the 
controls $\Delta{}^e\bm{m}_s$ have the properties given in equation~\eqref{eq:change_of_variables}. At this stage we 
orthogonalise 
and must relate derivatives with respect to $\bm{m}^n_s$ and $\tilde{\bm{m}}^n_s$ by the following
(valid for time level~$n$):
\begin{equation}\label{partial-F-update}
\widehat{\frac{{\rm d} F}{{\rm d} {\bm{m}}_s^{n} }}
= \frac{{\rm d} \tilde{\bm{m}}_s^{n} } {{\rm d} {\bm{m}}_s^{n} } \widehat{\frac{{\rm d} F}{ {\rm d} 
\tilde{\bm{m}}_s^{n} }}
=:({{\bm{V}}^{n}})^{-T} \widehat{\frac{{\rm d} F}{{\rm d} \tilde{\bm{m}}_s^{n} }}
\end{equation}


In order to calculate the sensitivity map, we need to be able to calculate~$\bm{V}$. To derive an expression we apply 
the chain rule to the derivative of the solution variables giving 
\begin{equation}
\frac{ {\rm d} \bm{\Psi}^n}{ {\rm d} {\bm{m}}^n_s} 
\frac{ {\rm d} {\bm{m}}^n_s}{ {\rm d} \tilde{\bm{m}}^n_s} 
=
\frac{ {\rm d} \bm{\Psi}^n}{{\rm d} {\bm{m}}^n_s} 
{\bm{V}}^n
=
\frac{ {\rm d} \bm{\Psi}^n}{ {\rm d} \tilde{\bm{m}}^n_s}\,.
\end{equation}
Combining this for all ensemble members and using the approximation $ {\bm{M}}_s^n \approx \widehat{\bm{M}}_s^n$, 
we assume that 
\begin{equation}
\widehat{\bm{M}}_s^n {\bm{V}}^n = \widetilde{\bm{M}}_s^n.  
\end{equation}
With the identity $(\widetilde{\bm{M}}_s^n)^T \widetilde{\bm{M}}_s^n = \bm{I}$ in mind, we pre-multiply the above 
matrix equation by $(\widetilde{\bm{M}}_s^n)^T$ to give
\begin{equation}
\bm{V}^n = \left( (\widetilde{\bm{M}}_s^n)^T \widehat{\bm{M}}_s^n \right)^{-1}, 
\end{equation}
and thus:
\begin{equation}
({\bm{V}}^n)^{-T} = \left( (\widetilde{\bm{M}}_s^n)^T \widehat{\bm{M}}_s^n\right)^{T}. 
\end{equation}

So in a practical algorithm one simply updates $\widehat{\frac{ {\rm d} F}{ {\rm d} {\bm{m}}_s^n}}$ using 
equations~\eqref{eq:magic} and~\eqref{partial-F-update} working backwards through time from the last time level:
\begin{equation}\label{dfdm---}
{\bm{m}}_s^{n+1} 
\widehat{\frac{ {\rm d} F}{ {\rm d} {\bm{m}}_s^n }}
=
\left(\bm{V}^n\right)^{-T} \widehat{\frac{ {\rm d} F}{ {\rm d}
\tilde{\bm{m}}_s^{n} }}
=
\left(\bm{V}^n\right)^{-T} \widehat{\frac{ {\rm d} F}{ {\rm d} {\bm{m}}_s^{n+1} 
}},
\end{equation}
and then applies equation~\eqref{eq:g_n} to obtain the sensitivity map $\widehat{\bm{g}}^n$ 
at time level~$n$.

\subsection{Typical algorithm for ensemble sensitivity calculation for time dependent problems}
\algrenewcommand\textproc{}
A typical algorithm for the goal-based approach outlined in section~\ref{sec:orthog_ensemble_members_biasing} 
can be seen in Algorithm~1. Extending this to include time windows (section~\ref{sec:time_windows}) is done in 
Algorithm~2. Both of these algorithms use two functions, \textproc{SolveEnsembles} and 
\textproc{CalculateSensitivityMap} described in Algorithms~3 and~4 
respectively. All other functions are not described as their form is implied by the function name. The 
function~\textproc{SolveEnsembles} inlcudes the smoothing, weighting and orthogonalising of the perturbations on line~3. 
The function~\textproc{CalculateSensitivityMap} shows how the sensitivity map is calculated, using all ensemble members 
for whichever time levels are desired. The inputs and outputs of the functions are shown in mathematical notation in 
order to facilitate comparison with the equations. 

\begin{algorithm}[htbp]
\caption{Calculate $\widehat{\bm{g}}$ using the goal-based approach 
(section~\ref{sec:orthog_ensemble_members_biasing}).}
\label{alg1_goal-based}
\begin{algorithmic}[1]
\State \Comment{Run unperturbed forward model}
\State $\overline{\bm{m}}$ = \Call{ReadInitialCondition}{\,}
\State $\overline{\bm{\Psi}}$ = \Call{RunForwardModel}{$\overline{\bm{m}}$}
\State $\overline{\bm{F}}$ = \Call{CalculateF}{$\overline{\bm{\Psi}}$}
\State
\State \Comment{Initialise sensitivity map vector}
\State $\widehat{\bm{g}}^0 = \bm{1}$ 
\State \Comment{Create the perturbations and solve the forward model}
\State $({}^1 \bm{\Psi}, {}^2\bm{\Psi}, \ldots, {}^{\cal{E}}\bm{\Psi}), ({}^1F, {}^2F, 
\ldots, {}^{\cal{E}} F ) $=\Call{SolveEnsembles}{$\widehat{\bm{g}}^0$, $\overline{\bm{m}}$}
\State
\State \Comment{Calculate the derivative of $F$ with respect to the controls, equation~\eqref{eqn:approxF}}
\State $\widehat{\frac{\rm{d}F}{\rm{d} \bm{m}_s}}$ = \Call{Calculate\,dFdms}{${}^1F, {}^2F, \ldots, {}^{\cal{E}}F, 
\overline{F}$}
\State
\State \Comment{Calculate the sensitivity map at each desired time level based on all $\cal{E}$ ensemble members, 
equation~\eqref{eq:g_n}}
\State $(\widehat{\bm{g}}^0, \widehat{\bm{g}}^1, \ldots, 
\widehat{\bm{g}}^n)$=\Call{CalculateSensitivityMap}{$({}^1\bm{\Psi}, {}^2\bm{\Psi}, \ldots, 
{}^{\cal{E}}\bm{\Psi}), ({}^1F, {}^2F, \ldots, {}^{\cal{E}}F )$, $\widehat{\frac{\rm{d} F}{\rm{d} 
\bm{m}_s}}$, $\overline{\bm{\Psi}}$, $\overline{F}$}
\end{algorithmic}
\end{algorithm}

\begin{algorithm}[htbp]
\caption{Calculate $\widehat{\bm{g}}$ using the goal-based approach and time windows 
(sections~\ref{sec:orthog_ensemble_members_biasing} and~\ref{sec:time_windows})
}
\label{alg2}
\begin{algorithmic}[1]
\State \Comment{Run unperturbed forward model}
\State $\overline{\bm{m}}$ = \Call{ReadInitialCondition}{\,}
\State $\overline{\bm{\Psi}}$ = \Call{RunForwardModel}{$\overline{\bm{m}}$}
\State $\overline{\bm{F}}$ = \Call{CalculateF}{$\overline{\bm{\Psi}}$}
\State
\State \Comment{Initialise sensitivity map vector}
\State $\widehat{\bm{g}}^{\text{init}} = \bm{1}$
\For {window $w = \,1$ to $W$ in reverse}
\State $[{}^1\bm{\Psi}, {}^2\bm{\Psi}, \ldots, {}^{\cal{E}}\bm{\Psi}]^w, [{}^1F, {}^2F, 
\ldots, {}^{\cal{E}}F]^w$ = \Call{SolveEnsembles}{$[\widehat{\bm{g}}^{\text{init}}]^w, \overline{\bm{m}}$}
\State
\State \Comment{Calculate the derivative of $F$ with respect to the controls using \ldots }
\If{$w == W$}\vspace{2mm}
    \State \Comment{\ldots equation~\eqref{eqn:approxF}}\vspace{1mm}
    \State$\left[\widehat{\frac{\rm{d} F}{\rm{d} \bm{m}_s}}\right]^w$ = \Call{Calculate\,dFdms}{$\left[{}^1F, 
{}^2F, \ldots, {}^{\cal{E}}F\right]^w, \overline{F}$}
\Else 
    \State \Comment{\ldots equations~\eqref{widehat M} and~\eqref{ll}}
    \State $\widehat{\bm{M}}_s^n $= \Call{Calculate\,d$\Psi$dm}{${}^1\bm{\Psi}^n, {}^2\bm{\Psi}^n, 
\ldots, {}^{\cal{E}}\bm{\Psi}^n, \overline{\bm{\Psi}}$}\vspace{1mm}
    \State $\left[\widehat{\frac{\rm{d} F}{\rm{d} \bm{m}_s}}\right]^w = [\widehat{\bm{M}}_s]^w 
[\widehat{\bm{g}}]^{w+1}$\vspace{1mm}
\EndIf
\State 
\State \Comment{Calculate the sensitivity map based on all $\cal{E}$ ensemble members, 
equation~\eqref{g_for_time_windows-}}
\State $(\widehat{\bm{g}}^0, \widehat{\bm{g}}^1, \ldots, 
\widehat{\bm{g}}^n)$=\Call{CalculateSensitivityMap}{$({}^1\bm{\Psi}, {}^2\bm{\Psi}, \ldots, 
{}^{\cal{E}}\bm{\Psi})$, $({}^1F, {}^2F, \ldots, {}^{\cal{E}}F )$,$\widehat{\frac{\rm{d} F}{\rm{d} 
\bm{m}_s}}$,$\overline{\bm{\Psi}}$, $\overline{F}$}
\EndFor
\end{algorithmic}
\end{algorithm}

\begin{algorithm}[htbp]
\caption{Create the perturbations and solve the ensemble members}
\label{fun1_solve_ensembles}
\begin{algorithmic}[1]
\Function{SolveEnsembles}{$\bm{\widehat{g}^0}$, $\overline{\bm{m}}$}
\For {ensemble $e = \,1$ to $\cal{E}$}
\State \Comment{Create perturbations}
\State $\Delta{}^e\bm{m}$ = \Call{GetPerturbation}{$\widehat{\bm{g}}^0$} \Comment{Using 
equations~\eqref{eq:smoothing_S},  \eqref{othog}, \eqref{eq:rescale} and~\eqref{g-bias}}
\State ${}^e\bm{m}$ = $\overline{\bm{m}} + \Delta{}^e\bm{m}$
\State \Comment{Solve the forward model}
\State ${}^e\bm{\Psi}$ = \Call{RunForwardModel}{${}^e\bm{m}$}
\State ${}^eF$ = \Call{CalculateF}{${}^e\bm{\Psi}$}
\State
\State \Comment{Calculate sensitivity map for time level 0 based on ensemble members $1$~to~$e$}
\State $\widehat{\bm{M}}^0$ = \Call{Calculate\,d$\Psi$dm}{${}^1\bm{\Psi}^0, {}^2\bm{\Psi}^0, 
\ldots, {}^e\bm{\Psi}^0, \overline{\bm{\Psi}} $} \Comment{equation~\eqref{widehat M} }
\State $\widehat{\frac{\rm{d} F}{\rm{d} \bm{m}_s}}$ = \Call{Calculate\,dFdms}{${}^1 F, {}^2 F, \ldots, {}^eF, 
\overline{F}$} \Comment{equation~\eqref{eqn:approxF} }
\State $\widehat{\bm{g}}^0$ = \Call{CalculateSensitivityMapTimeLevel}{$\widehat{\bm{M}}^0,\widehat{\frac{\rm{d} 
F}{\rm{d} \bm{m}_s}}$} \Comment{equation~\eqref{eq:g_n} }
\EndFor
\State \Return $({}^1\bm{\Psi}, {}^2\bm{\Psi}, \ldots, {}^{\cal{E}}\bm{\Psi}), 
({}^1F, {}^2F, \ldots, {}^{\cal{E}} F )$
\EndFunction
\end{algorithmic}
\end{algorithm}

\begin{algorithm}[htbp]
\caption{Calculates the sensitivity map}
\label{fun2_calc_g}
\begin{algorithmic}[1]
\Function{CalculateSensitivityMap}{$({}^1\bm{\Psi}, {}^2\bm{\Psi}, \ldots, 
{}^{\cal{E}}\bm{\Psi}), 
({}^1F, {}^2F, \ldots, {}^{\cal{E}}F )$,  $\widehat{\frac{\rm{d} F}{\rm{d} \bm{m}_s}}$, 
$\overline{\bm{\Psi}}$, $\overline{F}$}
\For {each time level $n$}
\State $\widehat{\bm{M}}^n $= \Call{Calculate\,d$\Psi$dm}{${}^1\bm{\Psi}^n, {}^2\bm{\Psi}^n, 
\ldots, {}^{\cal{E}}\bm{\Psi}^n, \overline{\bm{\Psi}}$} \Comment{equation~\eqref{widehat M}}
\State $\widehat{\bm{g}}^n$ = \Call{CalculateSensitivityMapTimeLevel}{$\widehat{\bm{M}}^n,\widehat{\frac{\rm{d} 
F}{\rm{d}{m}}_s}$} \Comment{equation~\eqref{eq:g_n}}
\EndFor
\State \Return $(\widehat{\bm{g}}^0, \widehat{\bm{g}}^1, \ldots, \widehat{\bm{g}}^n)$
\EndFunction
\end{algorithmic}
\end{algorithm}

\section{Results}
\label{Results} 

Problems from different subject areas are tackled using different codes to prove the robustness of the presented method.
First, it is tested in a 1D code for advection (see section~\ref{1D_advection}). Next, the method is used to analyse a 
2D system in which we model the advection of a tracer (see 
section~\ref{2D_advection}) with the code Fluidity~\cite{Pain_2001,fluidity_manual, xie_2016} 
and using the flux limited advection method~\cite{gomes_2016}. This advection method 
uses, as the high order solution, a Finite Element Galerkin projection of the control volume 
solution with linear triangular elements.   
Finally, the formulation is used to study the behaviour of a non-linear multi-phase porous 
media flow in a 3D heterogeneous media (section~\ref{3D_porous}) using ICFERST~\cite{jackson_2015,gomes_2016, 
salinas_2017}.

For test cases \ref{1D_advection} and \ref{2D_advection} we solve the advection equation:
\begin{equation}
\frac{\partial c}{\partial t} +\bm{u}\cdot\nabla c = 0 \,,
\label{1d_adv_eqs}
\end{equation}
in which $\bm{u}=1$ for 1D, $\bm{u}=(1,0)$ for 2D, and where $c$ is the concentration of the tracer. 

%

For readability, a brief summary of the equations to solve for multi-phase porous media flow (test case \ref{3D_porous})
are presented. A more complete description of the formulation has been 
published~\cite{jackson_2015,gomes_2016,salinas_2017}.
Darcy's equation is as follows:
\begin{equation}
  \mu_\alpha S_\alpha \left(\mathcal{K}_{{r}_\alpha}\mathbf{K}\right)^{-1}
  \bm{u}_{\alpha} = - \nabla p +
         {\bm{s}_{u}}_{\alpha},
  \label{force-bal}
\end{equation}
in which  $\bm{u}_\alpha$ is the velocity of phase $\alpha$, $p$ is the global pressure of the system and 
$\bm{s}_{{u}_\alpha}$ is a source term; here no sources are considered. $\mathbf{K}$ is the permeability tensor, and 
$\mathcal{K}_{{r}_\alpha}$, $\mu_\alpha$ and $S_\alpha$ are the relative permeability, viscosity and saturation of 
phase $\alpha$ respectively. 

The saturation equation for incompressible flow is:
\begin{equation}
  \phi\displaystyle\frac{\partial S_{\alpha} }{\partial t} + \nabla
  \cdot \left( {\bm u}_{\alpha} S_{\alpha}\right) =0,
  \label{sat_eq_pe_stab}
\end{equation}
in which $\phi$ is the porosity. 

The system of equations is closed by ensuring that the phase saturations sum to one:
\begin{equation}
  \sum\limits_{\alpha=1}^{n} S_{\alpha} =
  1,
  \label{sum_to_one}
\end{equation}
$n$ being the number of phases. 
For the relative permeability, the Brooks-Corey model~\cite{brooks_1964} is used
\begin{eqnarray}
k_{rw}\left ( S_{w} \right ) &=& \left ( \frac{S_w-S_{wirr}}{1-S_{wirr}-S_{nwr}} \right )^{n_w}, \\
k_{rnw}\left ( S_{nw} \right ) &=& \left ( \frac{S_{nw}-S_{nwr}}{1-S_{wirr}-S_{nwr}} \right)^{n_{nw}}, 
\end{eqnarray}
in which $n_w$ and $n_{nw}$ are the exponents for the wetting and non-wetting phases respectively. $S_{nwr}$  is the 
irreducible non-wetting phase saturation and $S_{wirr}$ is the irreducible wetting phase saturation.

\subsection{1D advection equation test}
\label{1D_advection} 
Equation~\eqref{1d_adv_eqs} is solved with an initial condition of $c=0$. 
The number of cells is set to ${\cal N}=101$ and the domain size is $100$ with the cell spacing equal to $\Delta x=1$. 
The time-step size is set to $\Delta t = 0.1$, the default number of equally-sized time steps is ${\cal N}_t =600$ %
and a forward Euler scheme is used. In this 1D test, the functional that we use in forming the sensitivities to the 
solution variables ($c_i^n$) is 
\begin{eqnarray}
F = c_k^{{\cal N}_t} \;\; {\rm with}\;\; k={ \left\lfloor{ 0.85 {\cal N} }\right\rfloor }. 
\end{eqnarray}
That is, the concentration at the final time level ${{\cal N}_t}$ and at cell $\left\lfloor{ 0.85 {\cal N} 
}\right\rfloor$ which is the nearest integer from below to $0.85 {\cal N}$. The cells are ordered from left to right 
with increasing $x$-coordinate. Thus the exact sensitivities at the end of time $n={{\cal N}_t}$ are:  
\begin{eqnarray}
\frac{\rm{d} F}{\rm{d} c_i^{{\cal N}_t}} = 1, \;\; {\rm if} \;\; i=\left\lfloor{ 0.85 {\cal N} }\right\rfloor \;\; 
{\rm and} \;\; \frac{\rm{d} F}{\rm{d} c_i^{{\cal N}_t}} = 0 \;\; {\rm if} \;\; i\neq \left\lfloor{ 0.85 {\cal N} 
}\right\rfloor.
\label{c-deriv}
\end{eqnarray}

The number of smoothing iterations is set to~$25$. To further 
explore the effectiveness of the presented method we use, either, an upwind differencing scheme (linear) or 
a Normalised Variable Diagram~\cite{Leonard91} (NVD) flux-limiting differencing scheme (non-linear). 
The high order scheme within the NVD method is the diamond differencing scheme (a mid-point scheme) and the 
NVD is such that diamond differencing is used as much as possible while not violating the 
Total Variational Diminishing criteria. 

%

\begin{figure}
	\centering\includegraphics[width=0.7\textwidth]{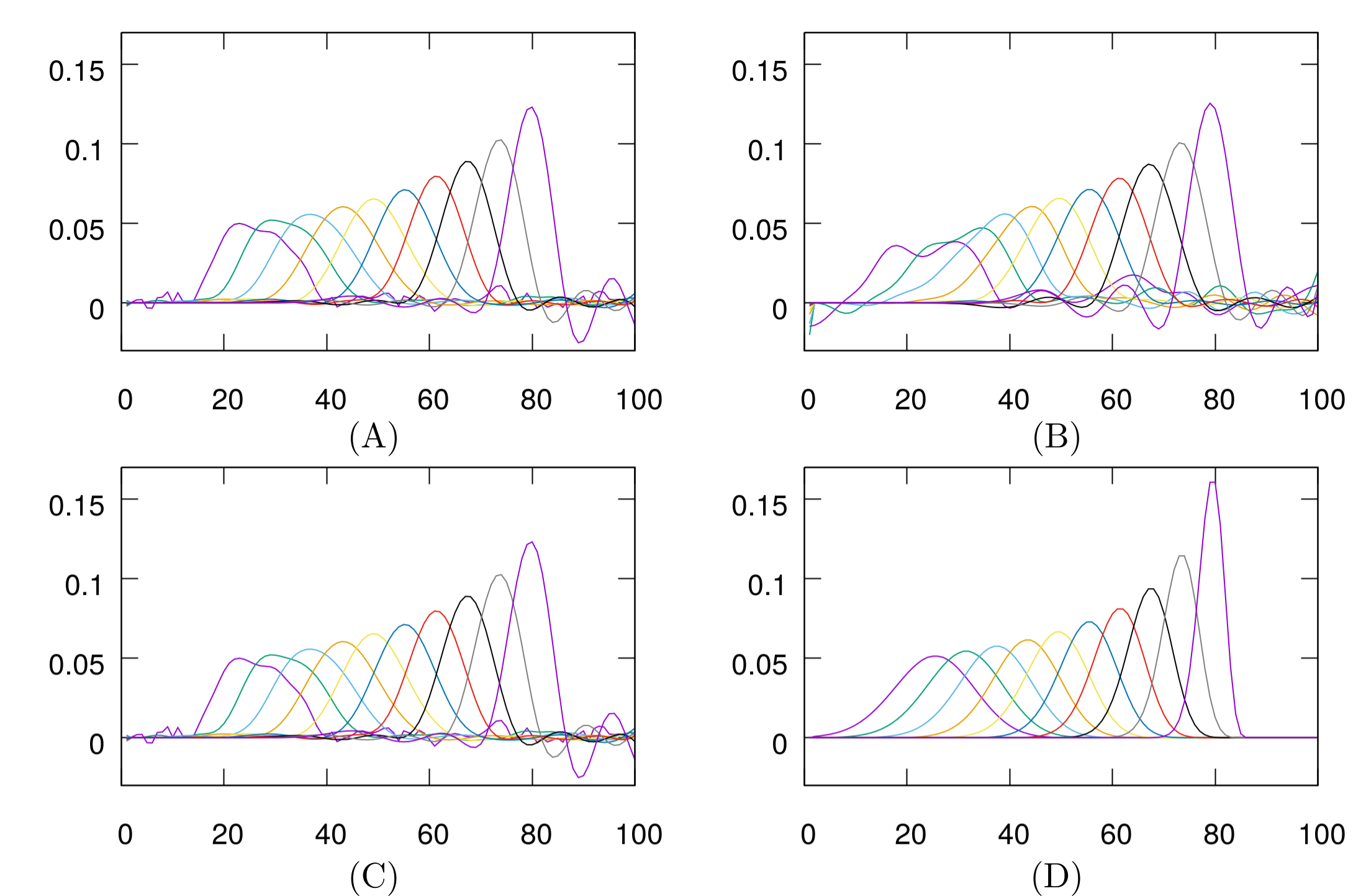}
\caption{The advection of sensitivities backwards through time using the upwind scheme. The $x$-axis shows the $x$ 
-coordinate, the $y$-axis 
shows the amplitude of the sensitivities.  Ten different time instances (represented by different curves) 
equally spaced from beginning to the end of the time are plotted.  (A) results using the goal-based approach, performing 
10 perturbations and without re-orthogonalisation.
(B) results without the goal-based approach, performing 10 perturbations and without re-orthogonalisation.
(C) results using the goal-based approach, performing 10 perturbations and with re-orthogonalisation.
(D) results using the goal-based approach, performing 101 perturbations and with re-orthogonalisation. 
 }
    \label{pics1}
\end{figure}

Figure~\ref{pics1} shows the results of advecting the sensitivities backwards through time. Upwind differencing is 
used in the forward model. Ten equally-spaced time instances (spanning the time domain) are plotted. Each 
curve represents a different time instance and displays the sensitivities ($\widehat{\bm{g}}^n$) at that time 
instance $n$. Hence, the peak of each curve occurs at a different position. Figure~\ref{pics1}~(A) displays the results 
using the goal-based approach, 10 perturbations and without re-orthogonalisation.  Figure~\ref{pics1}(B) displays 
the results using 10 perturbations without the goal-based approach and without re-orthogonalisation. Comparing 
these figures is effectively comparing the methods known in the literature, see section~\ref{Method of generating the 
ensembles}, with our goal-based approach, section~\ref{sec:orthog_ensemble_members_biasing}. It is clearly seen 
that without the goal-based approach, Figure~\ref{pics1}(B), there are more oscillations in the sensitivity map than 
with the goal-based approach, Figure~\ref{pics1}(A). 
Figure~\ref{pics1}(C) is obtained using the goal-based approach, 10 perturbations and re-orthogonalisation.
There is little difference between Figures~\ref{pics1}(C) and~\ref{pics1}(A), because, for such a low 
number of perturbations, the re-orthogonalisation has little effect as the perturbations naturally remain 
independent to one another. Finally, Figure~\ref{pics1}(D) displays the results using the goal-based 
approach, 101 perturbations and re-orthogonalisation. Equivalent to direct sensitivity analysis, the results 
for 101 perturbations are effectively the ``true'' numerical sensitivities. These sensitivities are in effect exact because, 
for the linear scheme, the resulting sensitivites are independent of the size of the perturbations. We have also varified the code and approach by directly perturbing each variable in order to produce the exact sensitivites which are identical to those shown in 
Figure~\ref{pics1}(D). It can be seen that 
with only 10 perturbations a good result can be obtained (compare Figure~\ref{pics1}(A) or (C) with 
Figure~\ref{pics1}(D)).

\begin{figure}
\centering\includegraphics[width=0.9\textwidth]{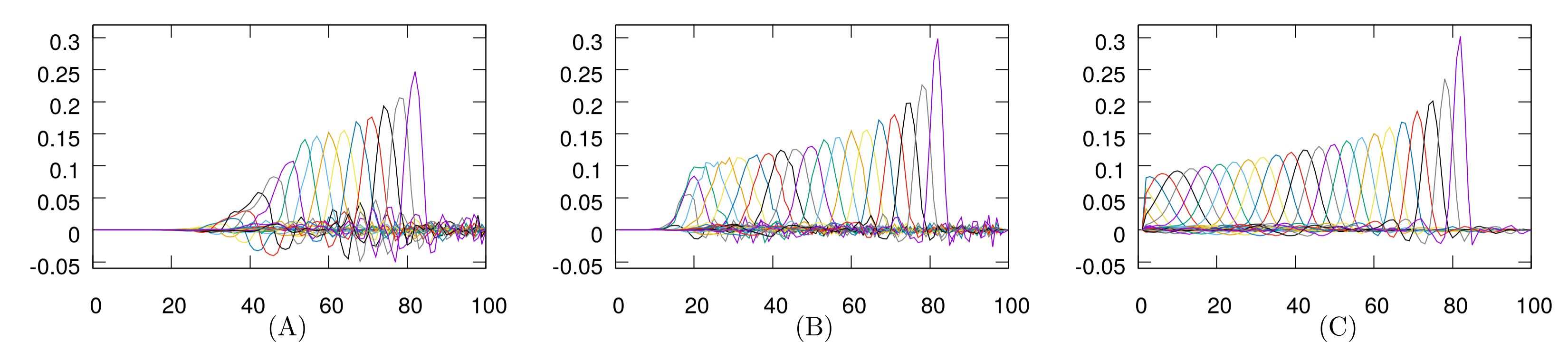}
\caption{Sensitivity results using the non-linear NVD flux-limiting scheme,  ${\cal N}_t=3600$ time steps with  
the goal-based approach and re-orthogonalisation. (A) results performing $20$ perturbations. (B) results 
performing $40$ perturbations. (C) results performing $101$ perturbations. The $x$-axis shows the $x$-coordinate, the 
$y$-axis shows the amplitude of the sensitivities.}
    \label{pics2}
\end{figure}

Figure~\ref{pics2} shows the results using the non-linear NVD flux-limiting scheme for ${\cal 
N}_t=3600$ time steps. The longer run time means that the information travels 3.6 times the length of the domain. Here, 
the goal-based approach and re-orthogonalisation are used in all three cases. This test demonstrates the 
benefits of re-orthogonalisation, as without it, the perturbations would be advected out of the domain 
therefore making it impossible to calculate a sensitivity map. The 
results are less dissipative than in the previous case due to the non-linear scheme used. 
Figure~\ref{pics2}~(C) displays the results using $101$ perturbations and represents the true numerical sensitivities  
through time. Figures~\ref{pics2}~(A) and~(B) show the results using $20$ and $40$ perturbations 
respectively. It can be seen how, in this case, with $20$ perturbations, the results are not as good as before with 
$10$. However, the results are good enough to show the sensitivities developing over the latter half of the numerical 
simulation. These results are much improved when using $40$ iterations, showing the sensitivities over the majority of 
the numerical domain.

\begin{figure}
\centering\includegraphics[width=0.7\textwidth]{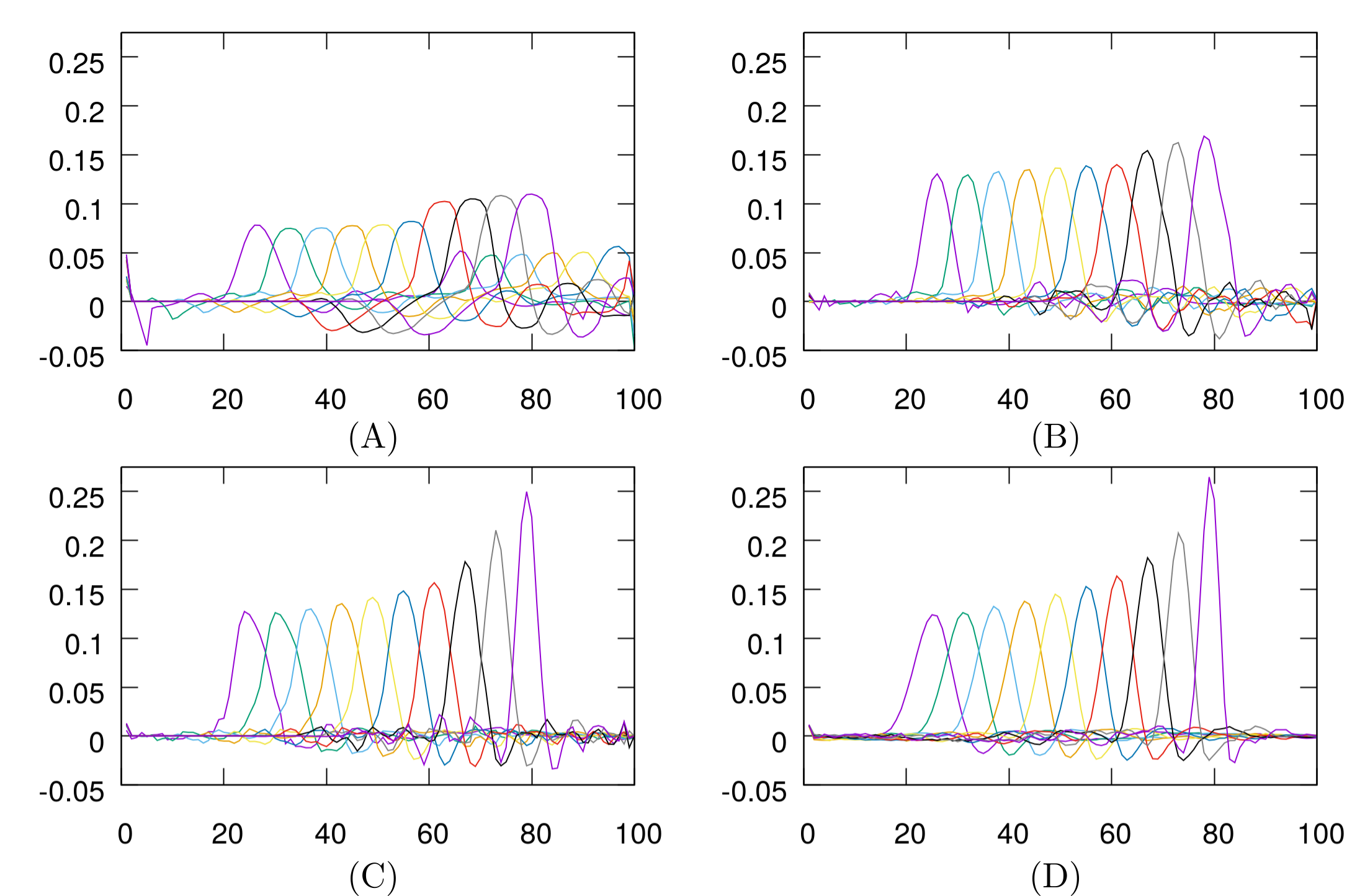}
\caption{
Sensitivity results using the non-linear NVD flux-limiting scheme, the goal-based approach and 
re-orthogonalisation. (A) results performing $5$ perturbations. (B) results performing $10$ perturbations. (C) 
results performing $20$ perturbations. (D) results performing $101$ perturbations. The $x$-axis shows the 
$x$-coordinate, the $y$-axis shows the amplitude of the sensitivities.}
    \label{pics3}
\end{figure}

Figure~\ref{pics3} shows the results of using the non-linear NVD flux-limiting scheme, the goal-based 
approach and re-orthogonalisation for 5, 10, 20 and 
101 perturbations in Figure~\ref{pics3}~(A), (B), (C) and~(D) respectively. Even with 5 
perturbations the sensitivities start to emerge and are well represented. With 10 
perturbations it can be seen that the results are improved. With 
20 perturbations the results are close to the final result obtained with 101 
perturbations, taken to be the true sensitivities.

\begin{figure}
\centering\includegraphics[width=0.7\textwidth]{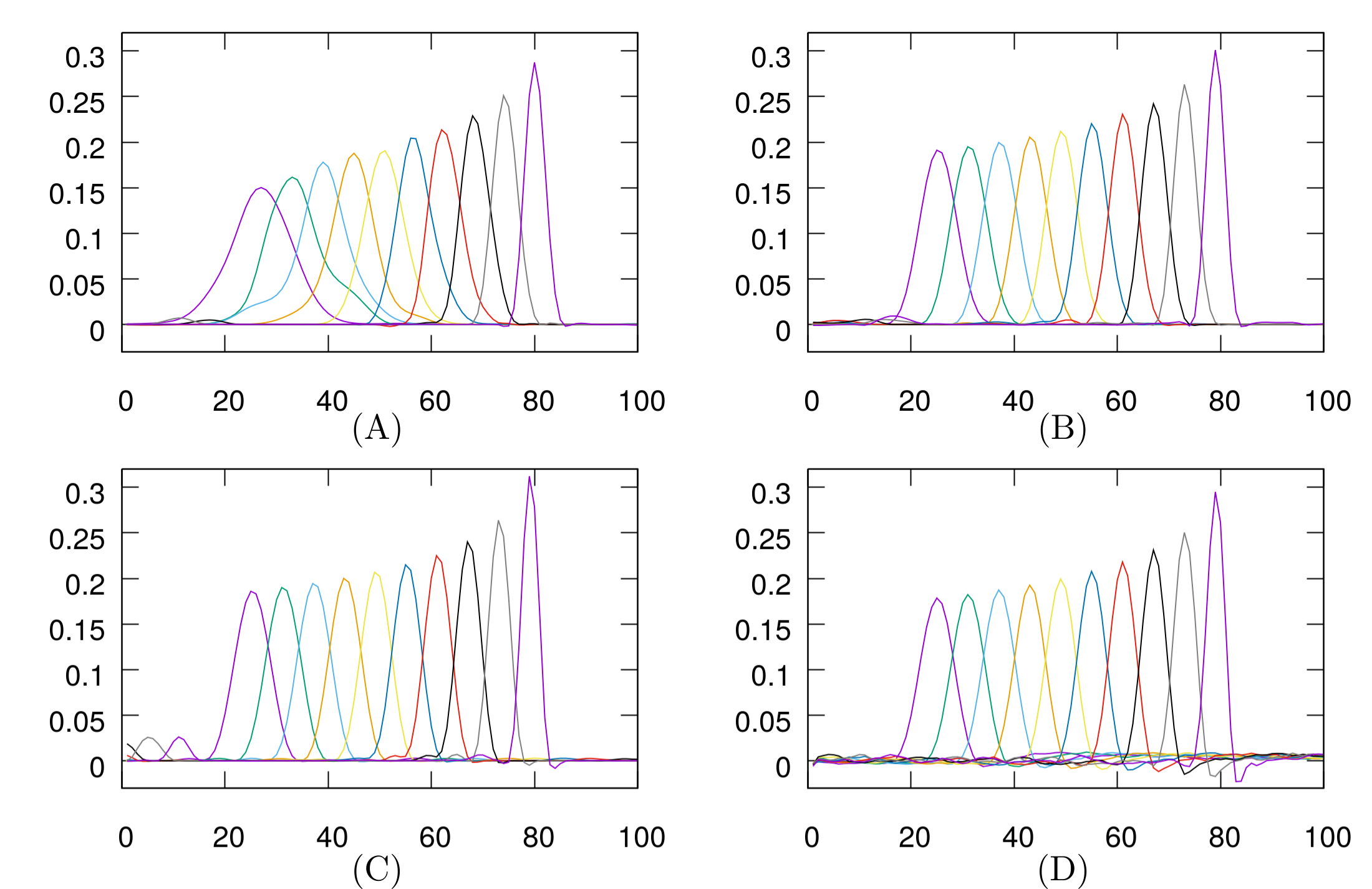}
\caption{
Sensitivity results using the non-linear NVD flux-limiting scheme, the goal-based approach and time windows. (A) results performing $5$ perturbations. (B) results performing $10$ perturbations. (C) results 
performing $20$ perturbations. (D) results performing $101$ perturbations. The $x$-axis shows the $x$-coordinate, the 
$y$-axis shows the amplitude of the sensitivities.}
    \label{pics3-1}
\end{figure}

In Figure~\ref{pics3-1} we repeat the test shown in Figure~\ref{pics3}{}, but now using time windows of 
size equal to one time step and working backwards through time. Since we are using time windows there is no need to use re-orthogonalization. Figures~\ref{pics3-1}~(A), (B), 
(C) and (D) have been obtained with 5, 10, 20 and 101 perturbations respectively. It can be seen that even with 5 
perturbations the results are excellent, representing the sensitivities well and minimising the oscillations.

\begin{figure}
\centering\includegraphics[width=0.8\textwidth]{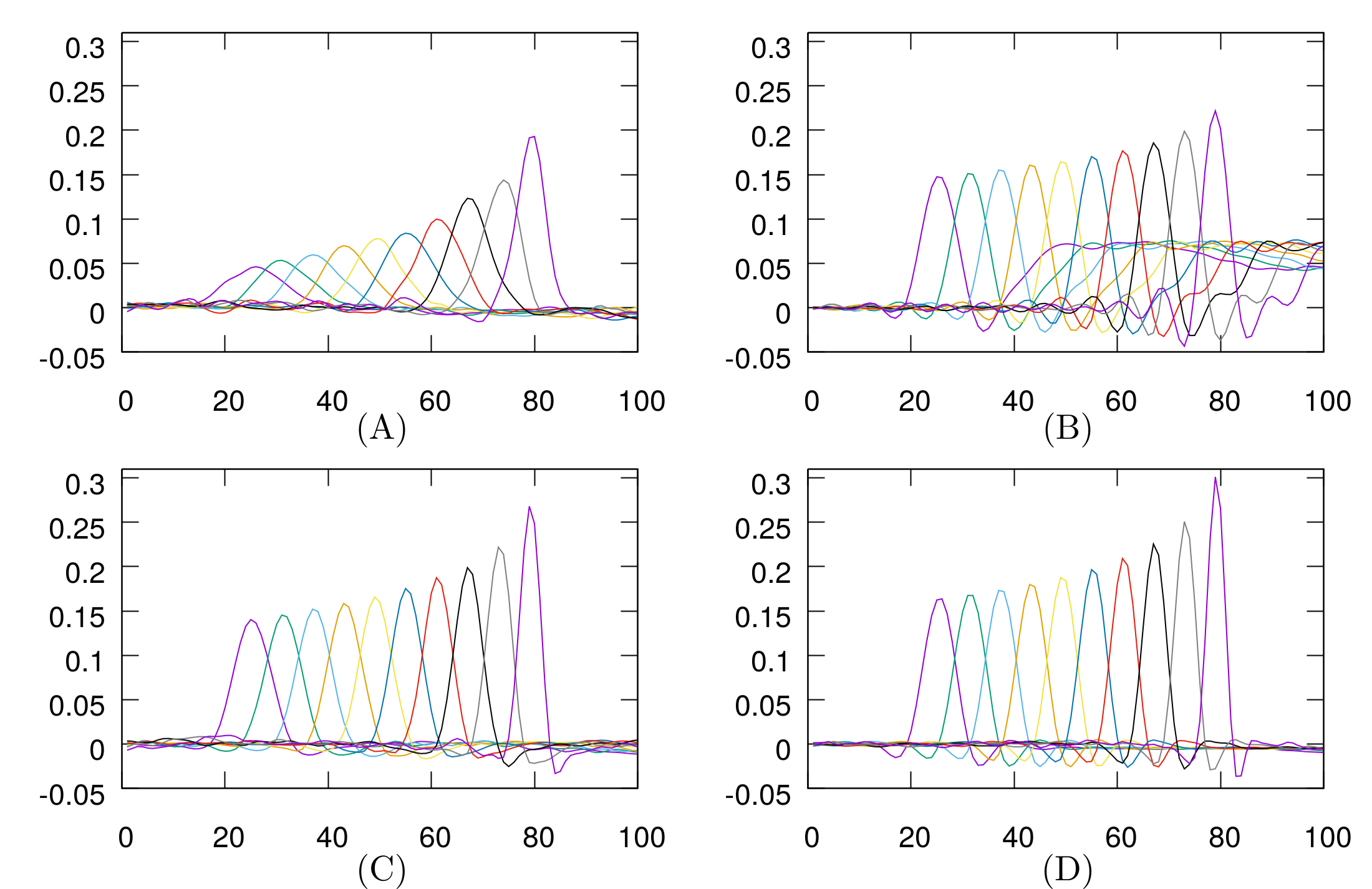}
\caption{
Sensitivity results using the non-linear NVD scheme, the goal-based approach (unless otherwise 
stated), re-orthogonalisation and explicit time windows. (A) results performing $20$ perturbations. (B) results 
performing $30$ perturbations without the goal-based approach. (C) results performing $30$ perturbations. (D) results 
performing $50$ perturbations. The $x$-axis shows the $x$-coordinate, the $y$-axis shows the amplitude of the 
sensitivities.} 
    \label{pics3-1-explicit}
\end{figure}

The test reported in Figures~\ref{pics3} and~\ref{pics3-1} is repeated once again using explicit time windows (that 
can be calculated concurrently), with the goal-based approach unless stated, for 25, 30 (without goal-based), 30 and 50 
perturbations and shown in Figure~\ref{pics3-1-explicit} (A), (B), (C) and (D) respectively. In this case, at least 25 
perturbations are required to obtain meaningful results. Moreover, for 30 
perturbations, not using the goal-based approach results in a much worse sensitivity map, seen by comparing 
Figure~\ref{pics3-1-explicit}~(B) with~(C). 
On comparing these results with those shown in Figure~\ref{pics3-1}{}, it can be seen that explicit time windows 
require  at least twice as many perturbations than standard time windows. However, explicit time windows do provide the 
opportunity to exploit parallelisation as all the perturbations and time windows can be performed at the same time. 
In this case 600 time windows are determined concurrently. Therefore, requiring more perturbations to obtain good 
results can easily be overcome by using the increased parallel capabilities introduced by this approach.


\begin{figure}
\centering\includegraphics[width=0.9\textwidth]{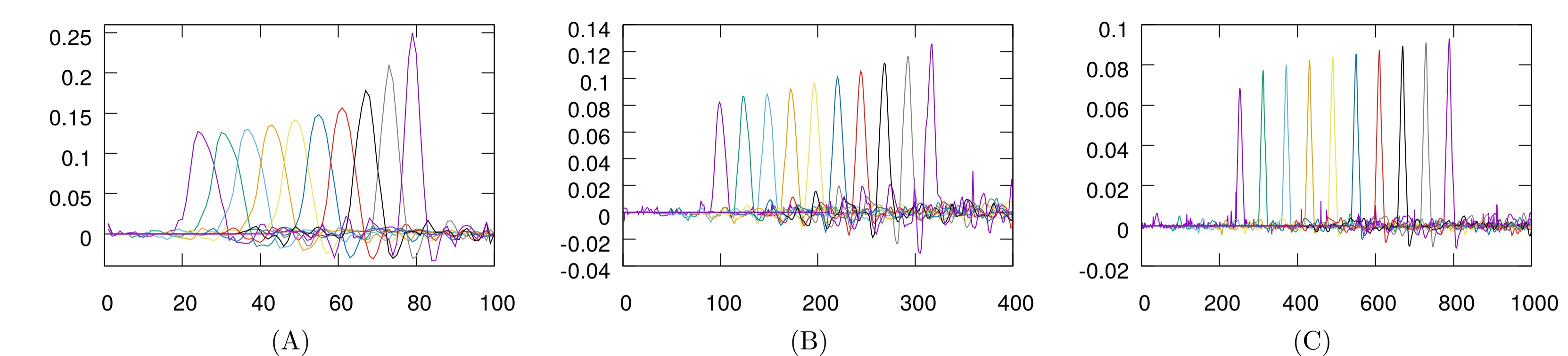}
\caption{Sensitivity results using the non-linear NVD scheme and the goal-based approach, where the initial perturbation 
is weighted by $\widehat{\bm{g}}^0,$ using only $20$ perturbations for three different meshes: $101$, $401$ and 
$1001$ cells, labelled (A), (B) and (C) respectively. The $x$-axis shows the $x$-coordinate, the $y$-axis 
shows the amplitude of the sensitivities.}
    \label{pics4}
\end{figure}

Figure~\ref{pics4} shows the results using the goal-based approach with only $20$ perturbations for different meshes 
($101$, $401$ and $1001$ cells). For each mesh, the time step is changed accordingly to ensure a constant Courant 
number. In all the cases, the results obtained are accurate. The sensitivities become sharper as the mesh is refined. 
We remark that resolving a numerical delta function is extremely demanding for ensemble methods, yet here, with just 20 
perturbations, the back-propagation of the sensitivities is well resolved even when using a very fine mesh 
(which increases the number of controls). 
It should be noted that the magnitude of the sensitivites decreases with increasing resolution. 
This is because, although the magnitude of the sensitivity maps begins with a magnitude of 1 in one cell (0 in all 
others), see equation~\eqref{c-deriv}, it quickly reduces (at least initially) as it spreads out and advects backwards 
in time.   


\begin{figure}

    \centering\includegraphics[width=0.6\textwidth]{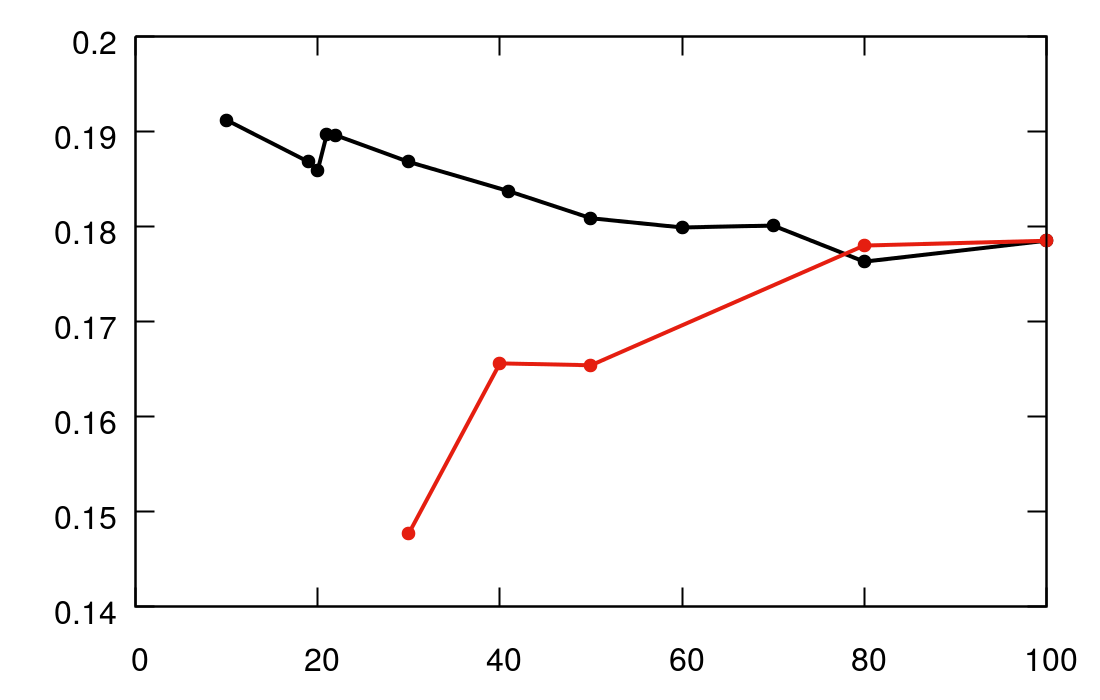}
\caption{The maximum value of $\widehat{\bm{g}}$ at $t=0$ is plotted against number of perturbations for results 
obtained using re-orthogonalisation, with (black) and without (red) $\widehat{\bm{g}}$ weighting the 
initial perturbations.  The number of cells is $101$ for this problem, and the non-linear NVD scheme is used. }
    \label{convergence_plot}
\end{figure}

From these numerical simulations it can be concluded that the goal-based approach is central to obtaining accurate 
sensitivity maps with fewer perturbations. Figure~\ref{convergence_plot} shows the convergence of the 
sensitivities for different numbers of perturbations, both with and without the goal-based approach. 
It should be noted that although the aim of some work is to have a degree of ensemble spread~\cite{Grimit_2007}, our 
aim 
is to explicitly have the most accurate sensitivities of the goal. 
Without the goal-based approach, results cannot be obtained unless a minimum of 30 perturbations are performed. 
Moreover, with 30 perturbations the results are worse than the case for the goal-based approach with 10 perturbations. 
As the number of perturbations is increased, the case using the goal-based approach consistently provides better 
results until reaching 80 perturbations (of a maximum of 101), when not using the goal-based approach is marginally 
better. 

\subsection{2D advection equation test}
\label{2D_advection}
\begin{figure}%
\centering
\includegraphics[width=6.5cm, trim = 50mm 170mm 10mm 20mm, clip]{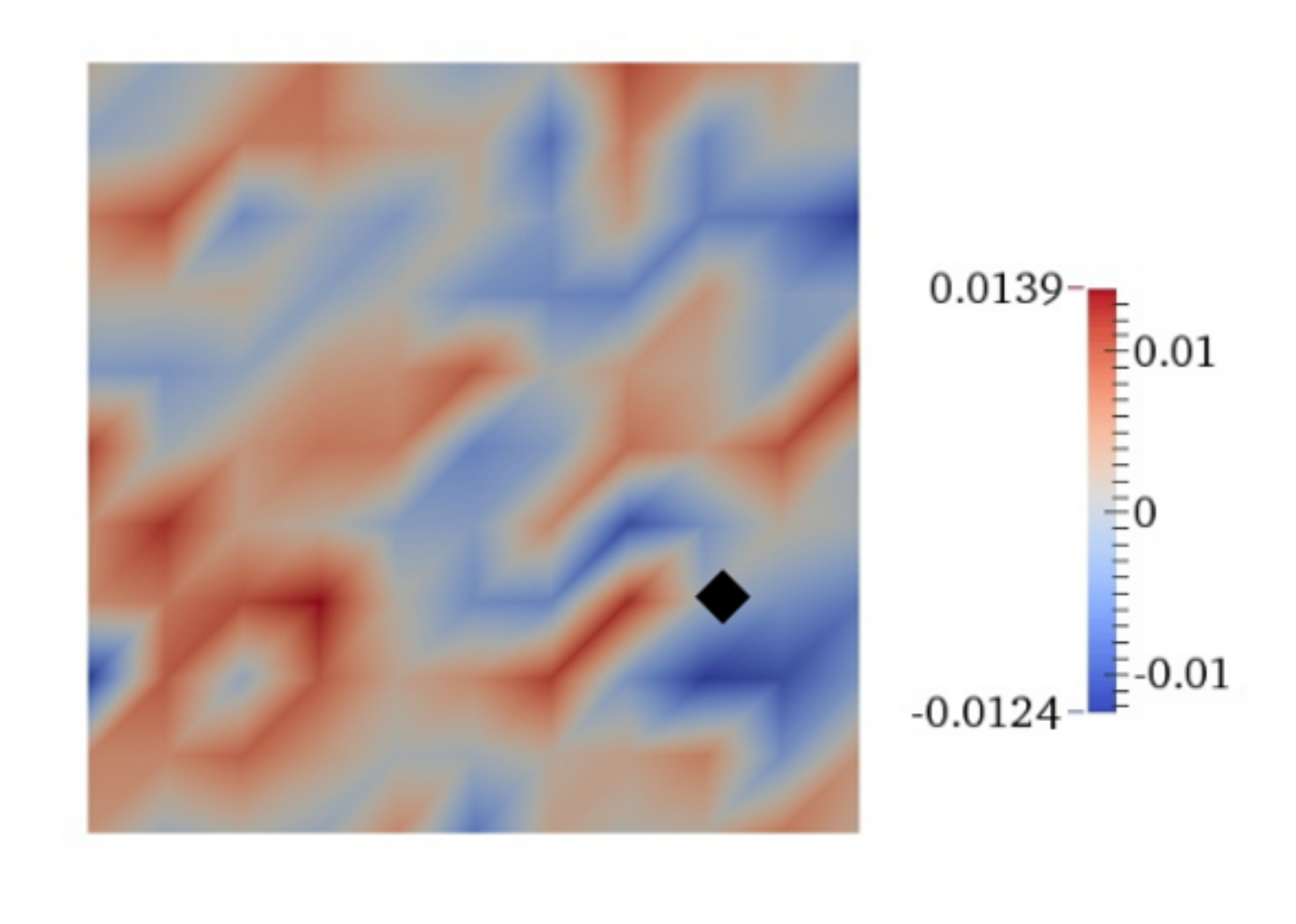}%
\includegraphics[width=6.5cm, trim = 50mm 170mm 10mm 20mm, clip]{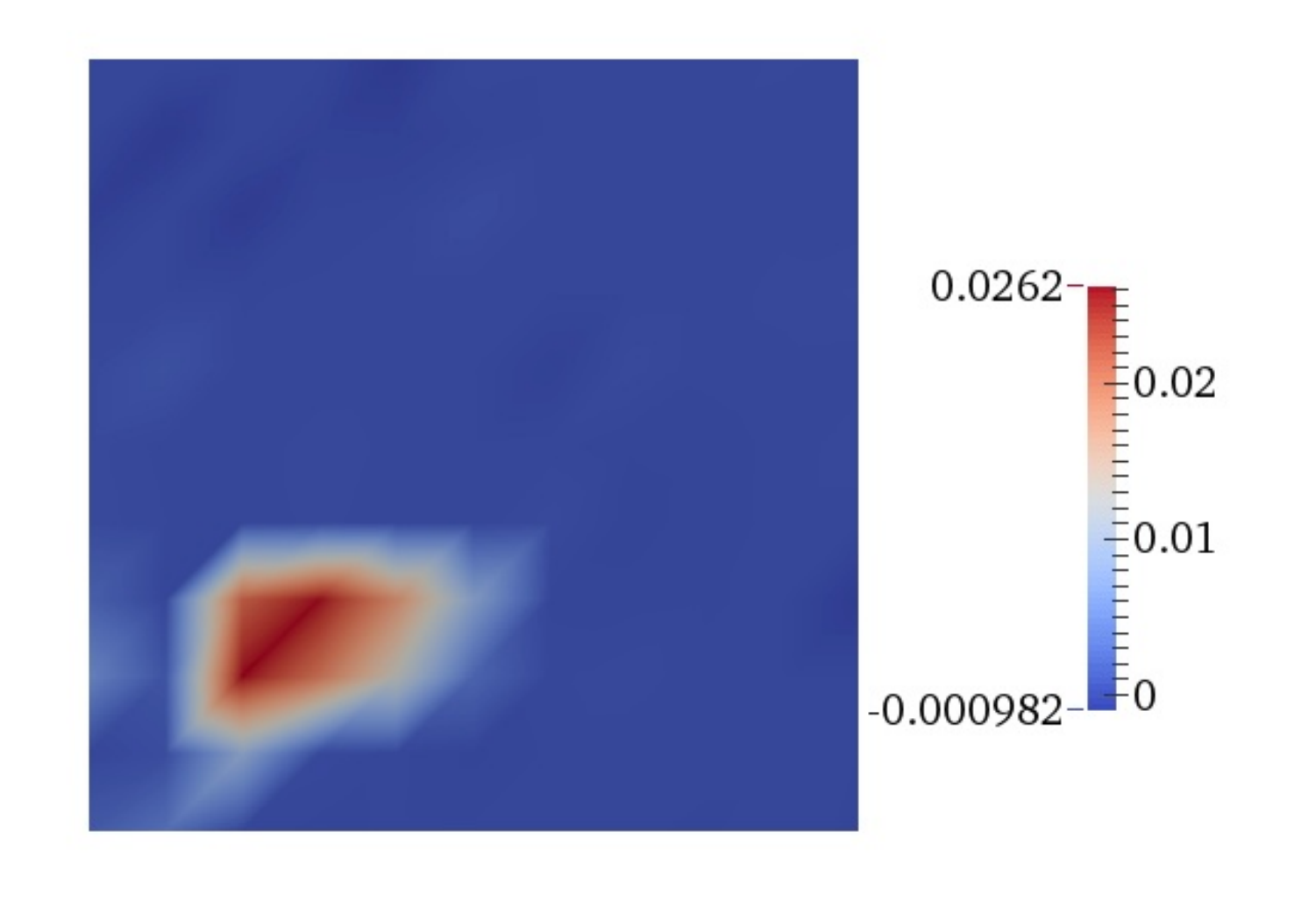}\\
\includegraphics[width=6.5cm, trim = 50mm 170mm 10mm 20mm, clip]{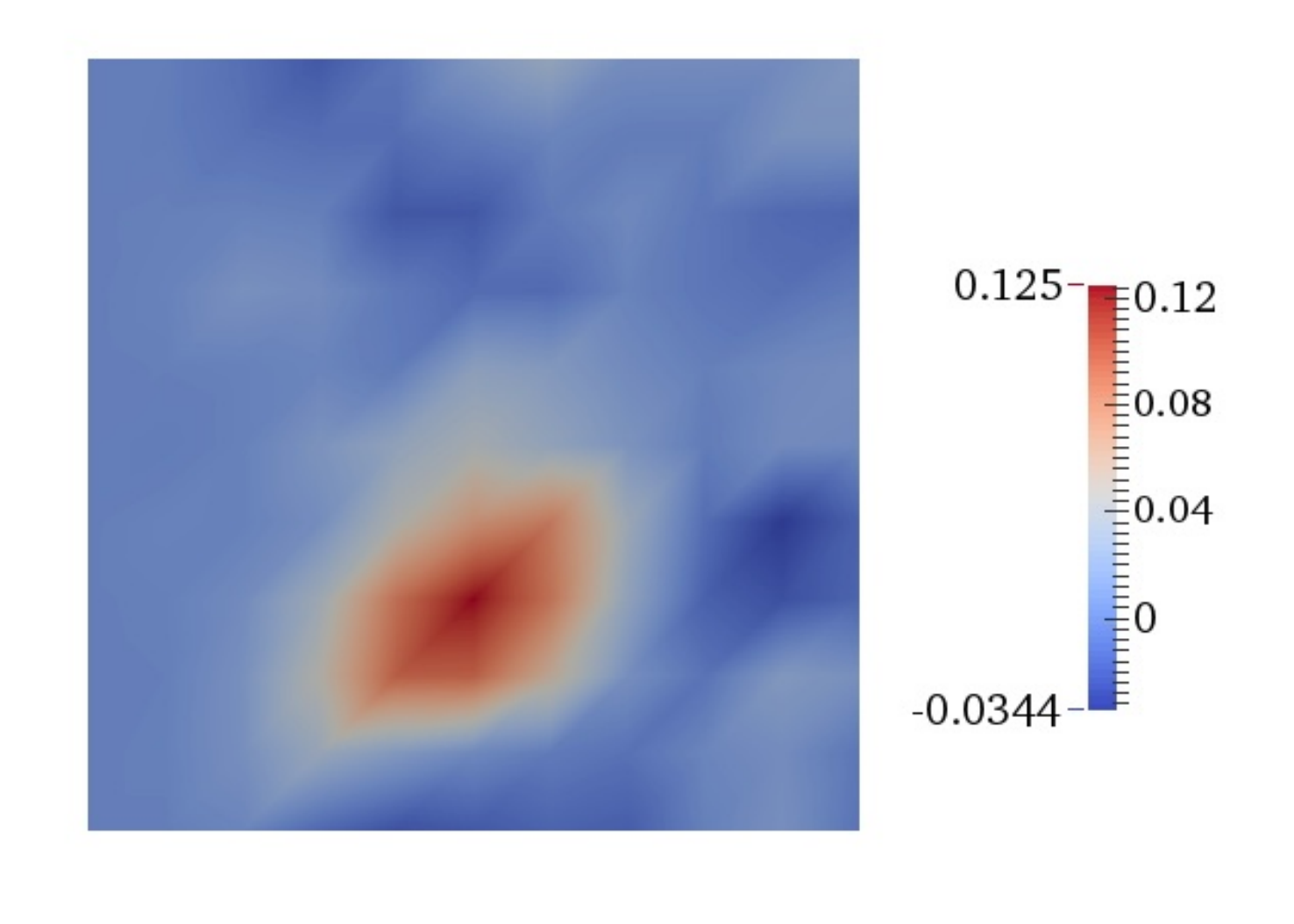}%
\includegraphics[width=6.5cm, trim = 50mm 170mm 10mm 20mm, clip]{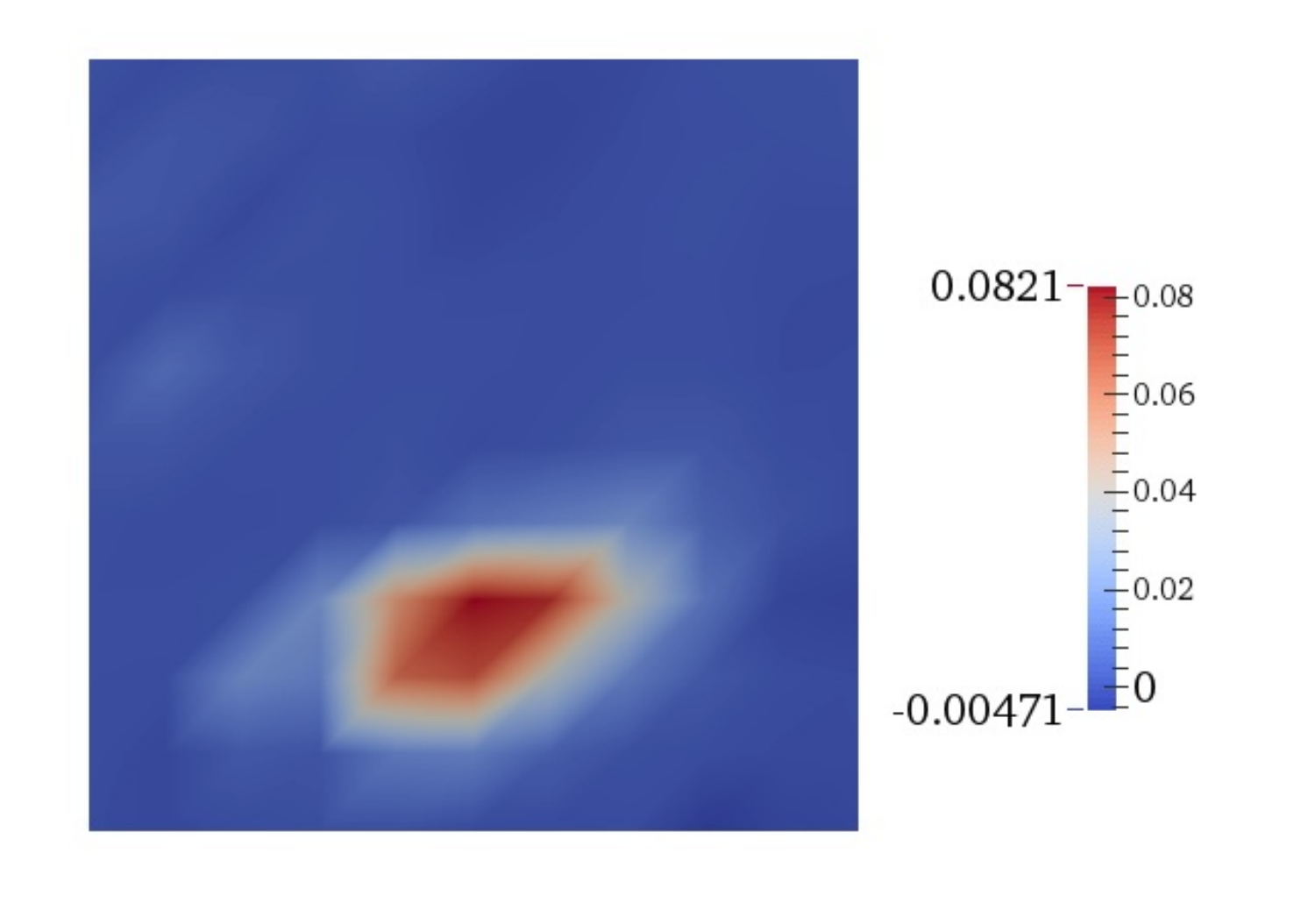}\\
\includegraphics[width=6.5cm, trim = 50mm 170mm 10mm 20mm, clip]{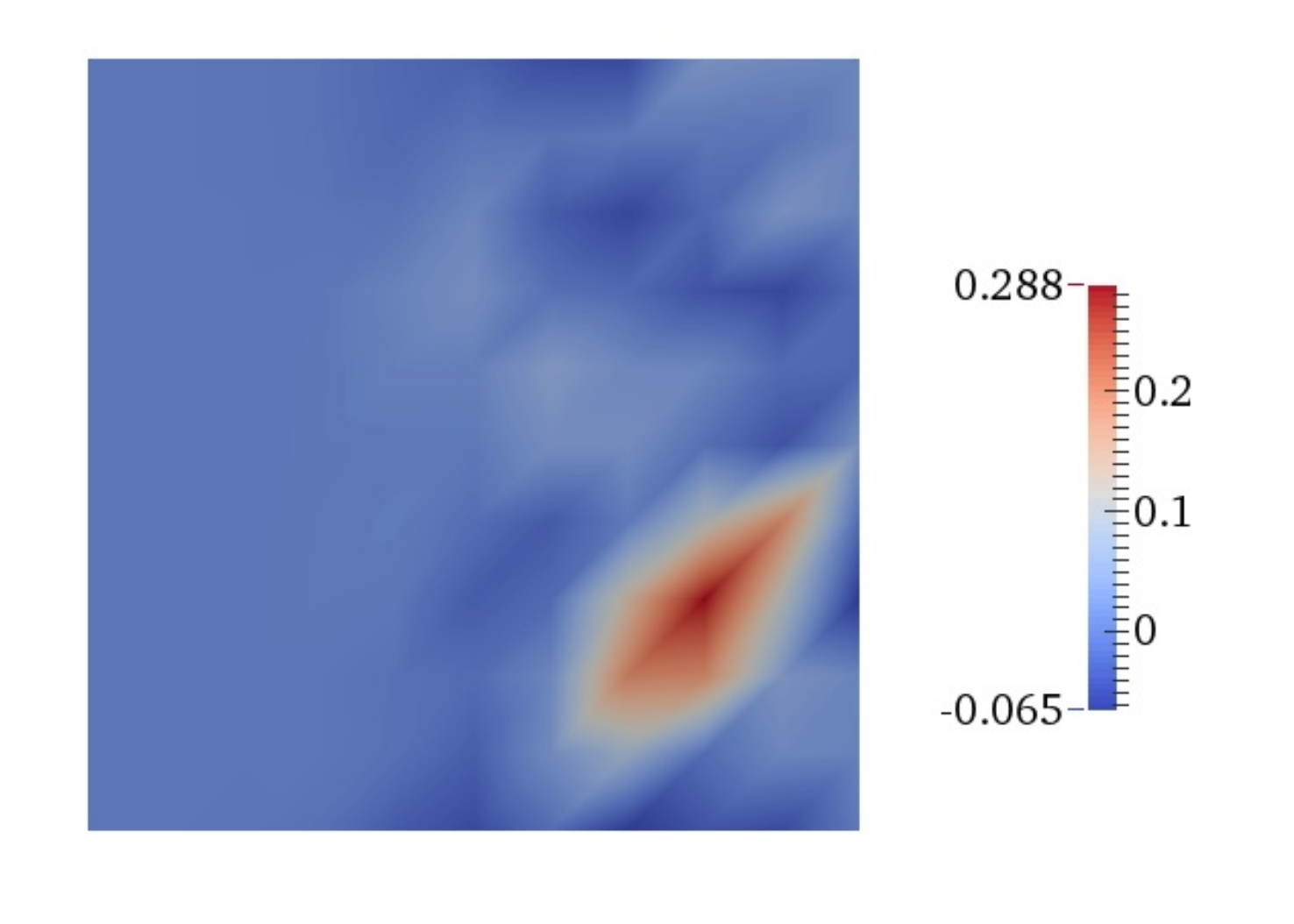}%
\includegraphics[width=6.5cm, trim = 50mm 170mm 10mm 20mm, clip]{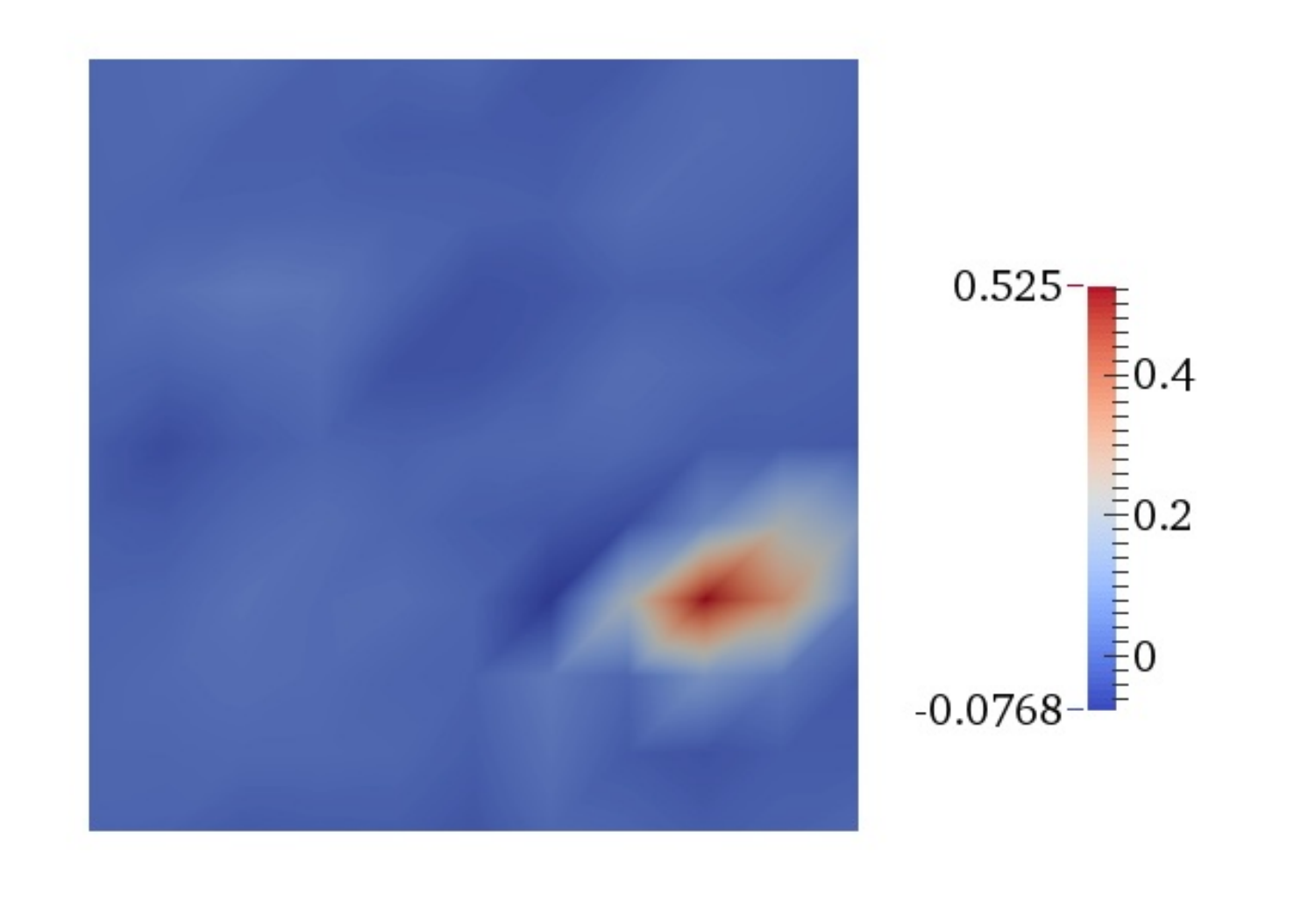}%
\caption{Results on the left show the sensitivity map generated with no smoothing, no weighting, 
no orthogonalisation and by using one time window. Results on the right are generated by smoothing the perturbations, 
weighting them with the sensitivity map available at that time, orthogonalising and by using seven time windows. The 
sensitivity map is shown at times $t=0$ (top), $t=1.75$ (middle) and $t=3.5$ (bottom). The size of ensemble is~20.}%
\label{fig:20pert_results}
\end{figure}

The governing equations for the 2D advection test are given in equation~\eqref{1d_adv_eqs}. For this test the velocity 
is set to $\bm{u} = (1,0)^T$, the domain size is 5 by 5 with a structured mesh of $100$ linear 
triangular elements and $121$ nodes (see Figure~\ref{fig:perturbation_steps}), the 
time step is set to 0.125 and the simulations are run from $t=0$ to $t=3.5$. The 
initial condition is $c(x,y,t=0) = 0.5$, and the boundary conditions are $c(0,y,t)=0.5$,  $c(5,y,t)=0$, $c(x,5,t)=0$ 
and $c(x,0,t)=0$. The functional in this problem is defined to be the concentration at the end of time 
($n={{\cal N}_t}$) at the point 
$(4,1.5)$ (indicated by a black diamond in Figure~\ref{fig:20pert_results} top-left). The analytical adjoint would be a 
dirac delta 
function which would move across the domain along the line $y=1.5$. The numerical solution suffers from dissipation, so 
the sensitivity map (i.e.~the numerical adjoint) will reflect this, spreading out and reducing in magnitude. 

Results are shown for one time window of 3.5 time units and seven time windows, each window of length 0.5 time units. 
For the one time-window case no special features are enabled (e.g.~smoothing, weighing, orthogonalisation etc.). For 
the seven time-window case, we weight the intial conditions with the sensitivity map available at the time, we 
orthogonalise 
and we pass down the value of the sensitivity map from one window to the previous. In both cases we use three 
smoothing iterations.
Figure~\ref{fig:20pert_results} shows results based on an ensemble size of~20. The plots in the left 
column pertain to one~time window, those on the right, to seven~time windows. The plots at the top are taken at the 
initial time $t=0$, those in the middle correspond to $t=1.75$ and those at the bottom are taken at the final time, 
$t=3.5$. The results for seven time windows show clearer sensitivity maps with less oscillation especially at earlier 
times. The peak value of the sensitivities is also larger for the seven window case. 

\begin{figure}%
\centering
\includegraphics[width=6.5cm, trim = 50mm 170mm 10mm 20mm, clip]{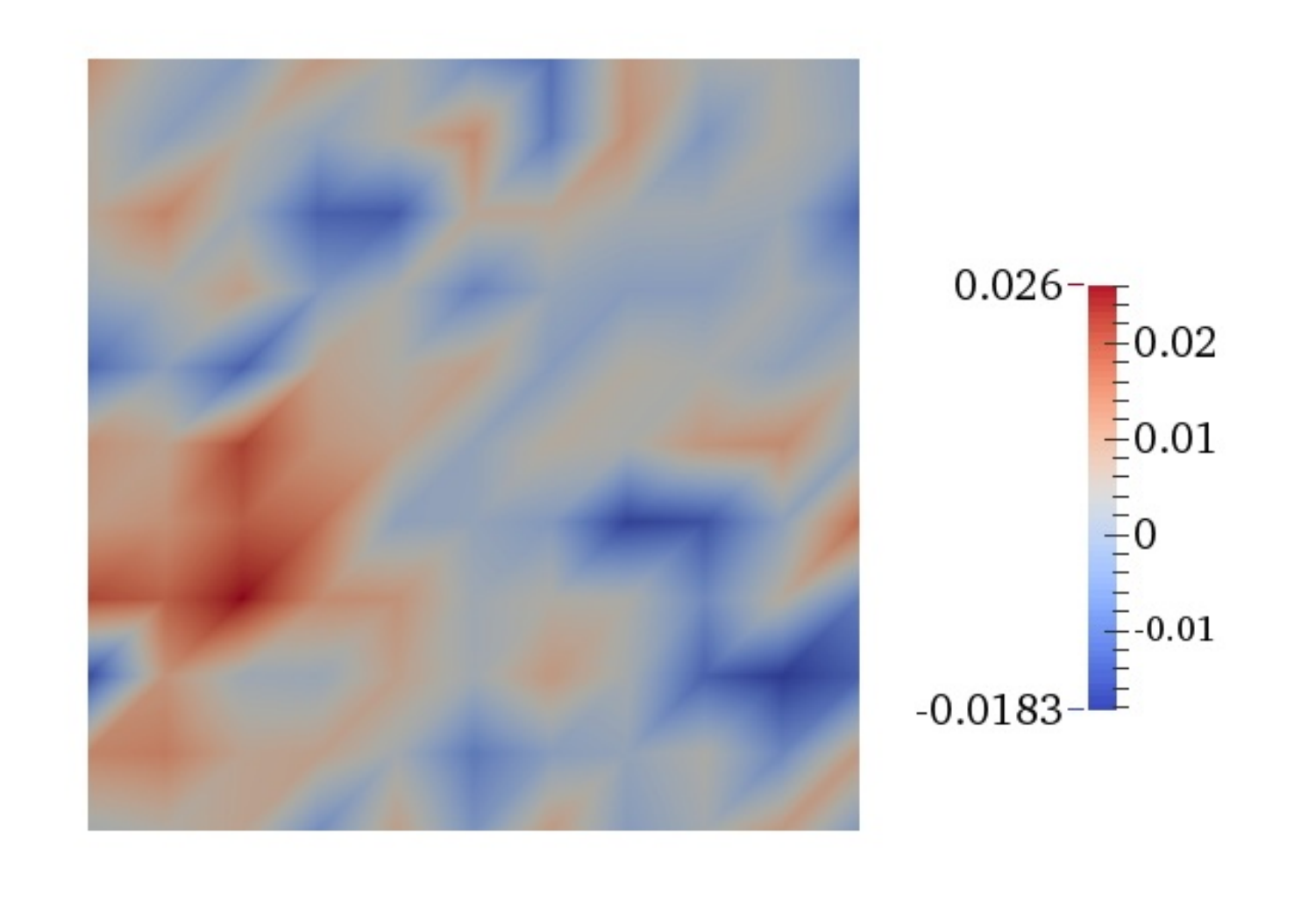}%
\includegraphics[width=6.5cm, trim = 50mm 170mm 10mm 20mm, clip]{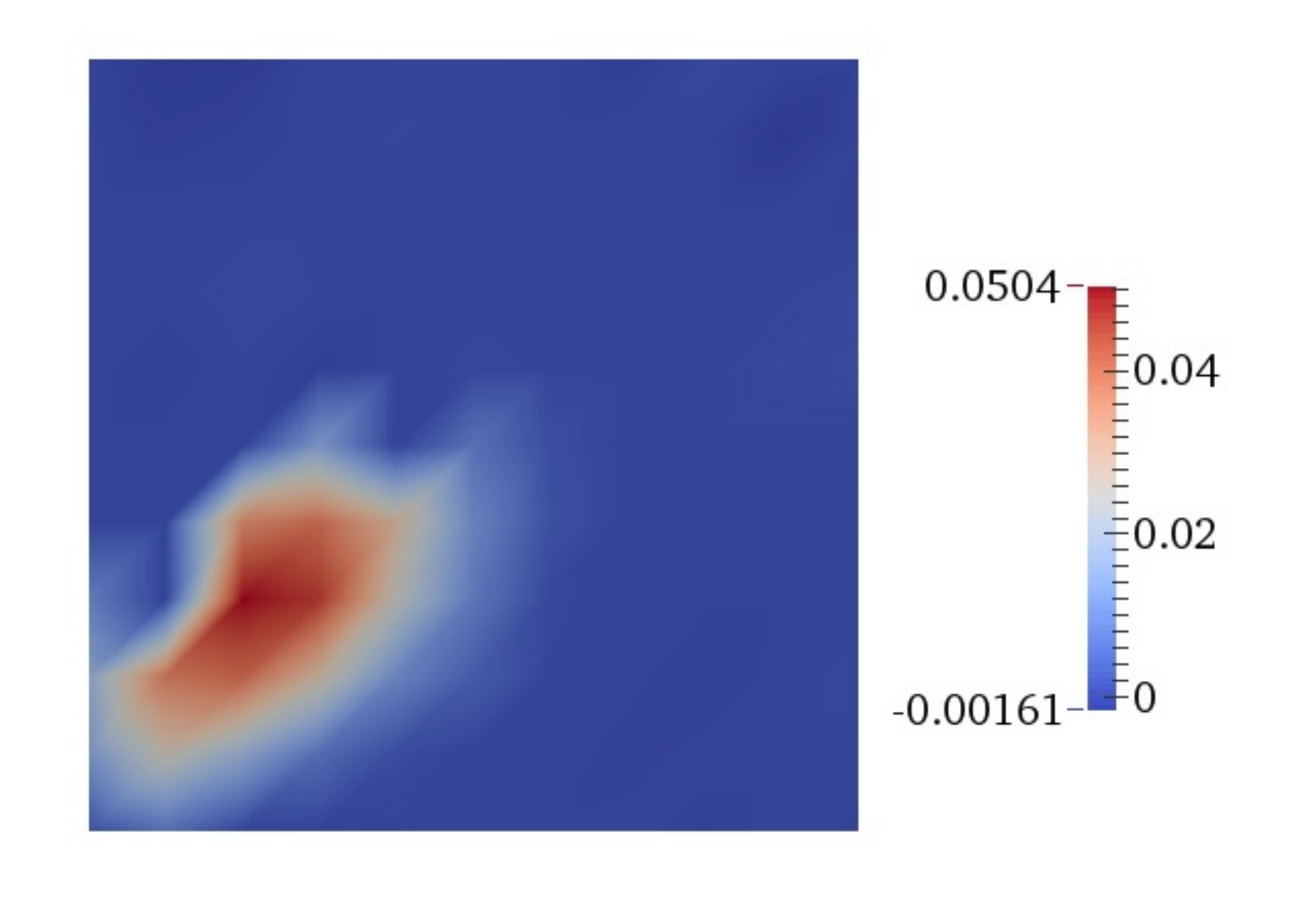}\\
\includegraphics[width=6.5cm, trim = 50mm 170mm 10mm 20mm, clip]{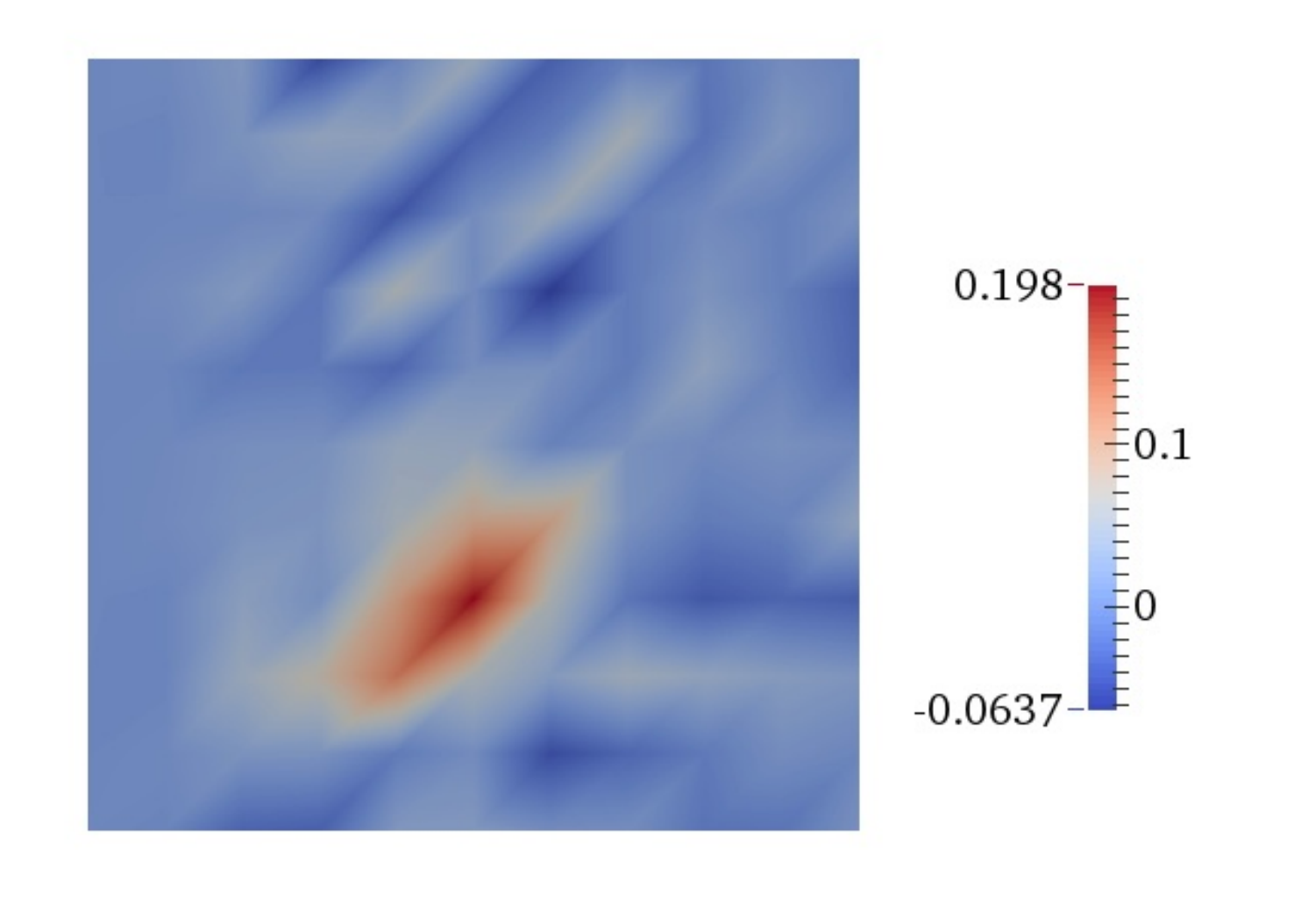}%
\includegraphics[width=6.5cm, trim = 50mm 170mm 10mm 20mm, clip]{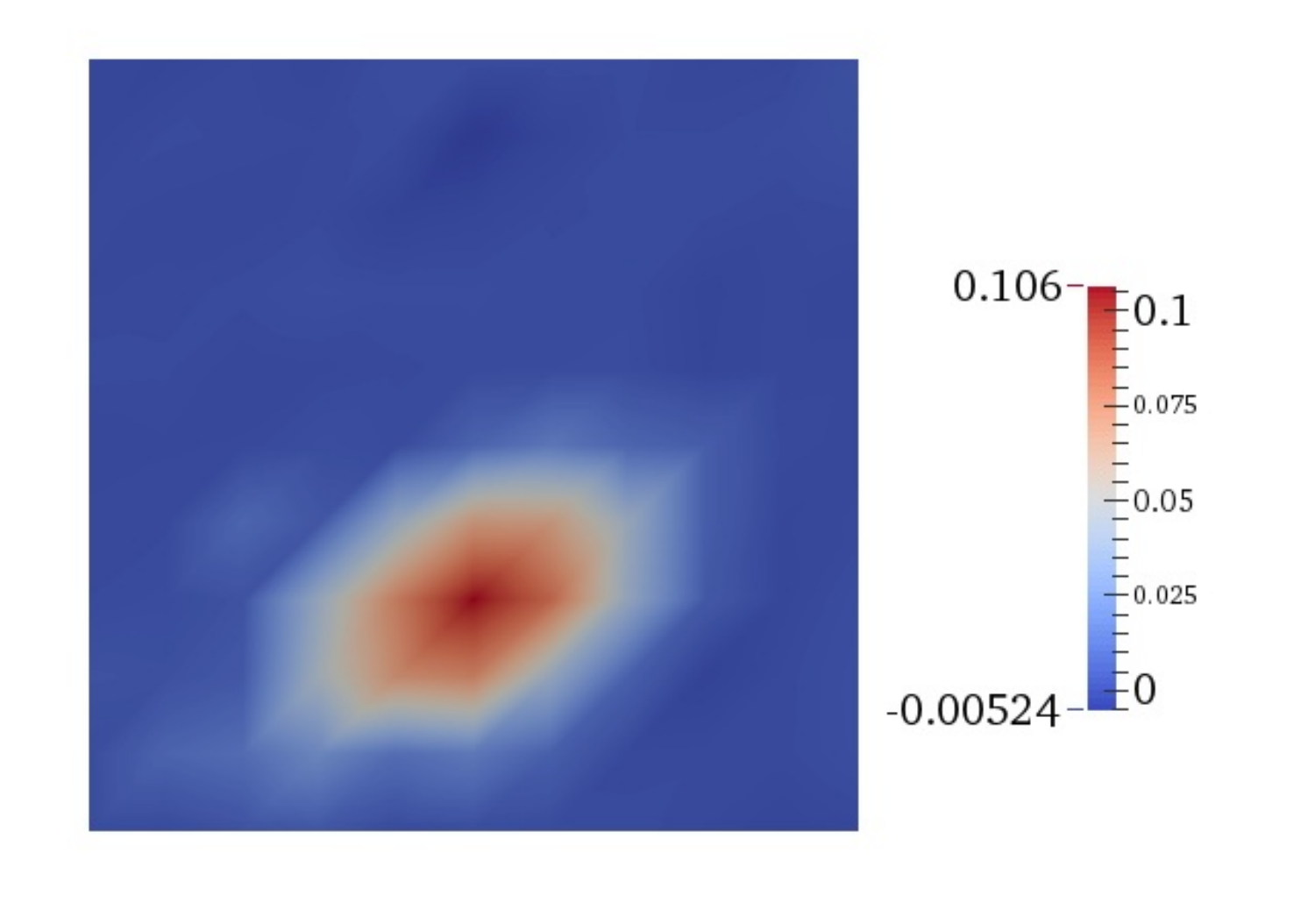}\\
\includegraphics[width=6.5cm, trim = 50mm 170mm 10mm 20mm, clip]{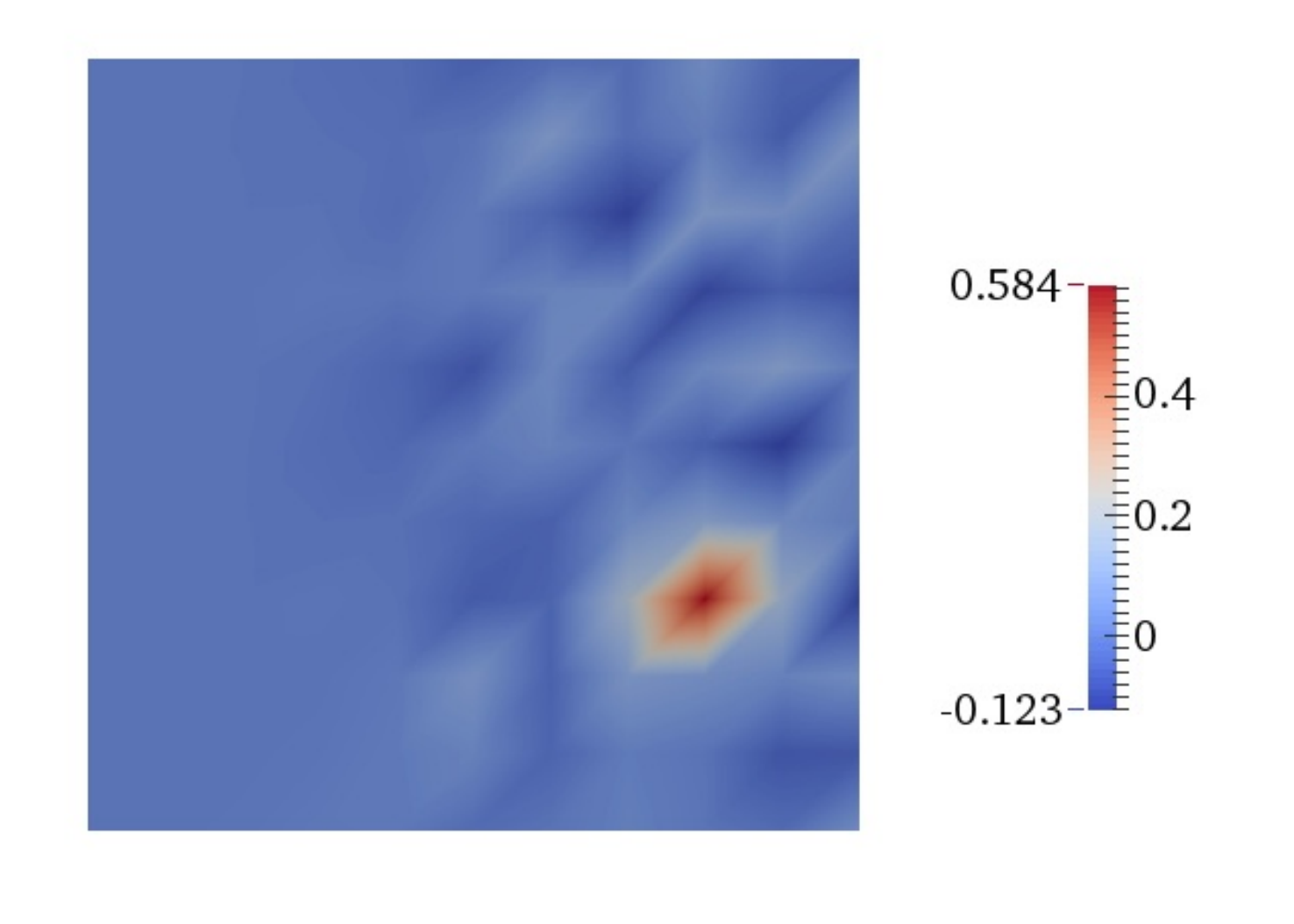}%
\includegraphics[width=6.5cm, trim = 50mm 170mm 10mm 20mm, clip]{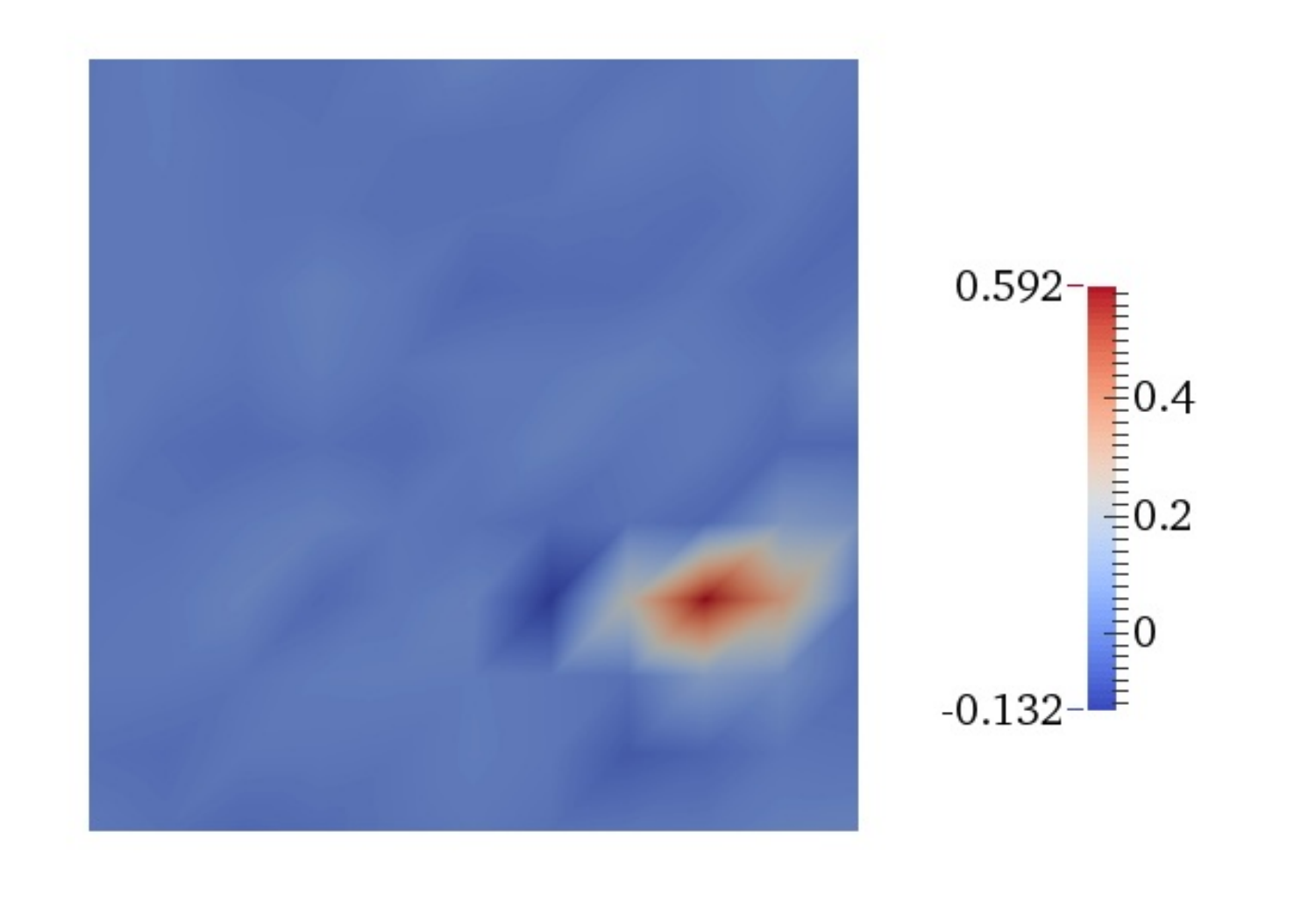}%
\caption{Results on the left show the sensitivity map generated with no smoothing, no weighting, no orthogonalisation 
and by using one time window. Results on the right are generated by smoothing the perturbations, weighting them with 
the sensitivity map available at that time, orthogonalising and by using seven time windows. The sensitivity map is 
shown at times $t=0$ (top), $t=1.75$ (middle) and $t=3.5$ (bottom). The size of ensemble is~40.}%
\label{fig:40pert_results}
\end{figure}

Figure~\ref{fig:40pert_results} shows results based on an ensemble size of~40. Again, the plots at the top are taken at 
the initial time $t=0$, those in the middle correspond to $t=1.75$ and those at the bottom are taken at the final time, 
$t=3.5$ with plots relating to one time window (and no additional features) on the left, seven time windows and all 
features on the right. Similar conclusions can be drawn to the previous case of 
20 ensemble members, i.e.~that the results for seven time windows show clearer sensitivity maps with less oscillation 
especially at earlier times, and the seven time-window case has a higher peak value. When using one time window and no 
additional features, the sensitivity maps are still not accurate even with an ensemble size of~40. 
When using seven time windows 
and all the features, an ensemble size of~20 may be enough (for some applications) to yield accurate 
sensitivities throughout time.

Comparing Figure~\ref{fig:20pert_results} (left) and Figure~\ref{fig:40pert_results} (left) shows the need to 
re-orthogonalise through time, as using larger ensemble sizes actually increases the noise seen in the sensitivity 
maps. This occurs due to the system becoming ill-posed, as time evolves, some random perturbations become similar to 
previously used perturbations. This can be solved either by introducing regularisation, by perturbing the boundary 
condition on the left of the domain or by ensuring that all the perturbations are orthogonal with respect to one 
another, which also increases the efficiency of the method.

%

\subsection{3D porous media test case} \label{3D_porous}
Here, the presented method is tested against a 3D multi-phase porous media flow. This test case has been chosen to 
explore the behaviour of the presented methodology against a highly non-linear system. The non-linearity arises from 
the relationship between the saturation and the relative permeability resulting in a non-linear correlation between the 
velocity of the phases and the saturation~\cite{jenny2009}. In this 
test case the saturation field is perturbed, as it affects the behaviour of the other fields. It is important to note 
that the amplitude of the perturbations performed is higher in this case.  
The range of the perturbations, in this case, is $0.1$, which is consistent with the perturbation amplitudes 
required to trigger viscous fingering~\cite{Jaure_2014}.

\begin{figure}[h!]
  \begin{center}
    \includegraphics[width=0.80\textwidth]{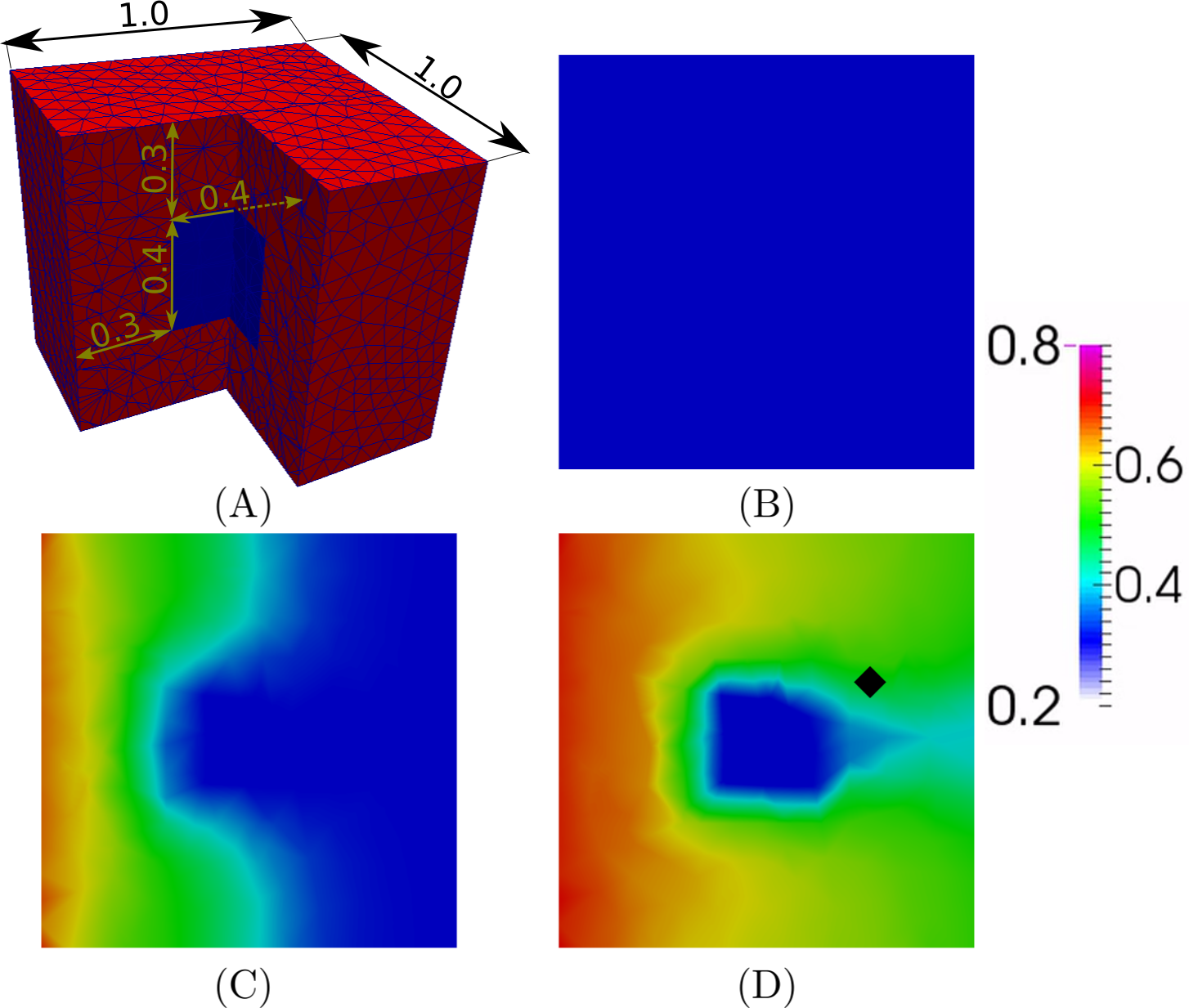}
    \caption{(A) Slice of the 3D domain showing the internal inclusion; domain dimensions, mesh (9879 elements) and 
permeability map are displayed. (B) (C) and (D) show the wetting saturation obtained from the forward model at 
three time instances. The diamond in (D) marks the point of interest used to create the sensitivity maps. 
\label{fig:Porous_forward}}
  \end{center}
\end{figure}

The porous medium consists of a box reservoir ($\mathbf{K}_{out} = 1$ dimensionless permeability units) with a low 
permeable inclusion ($\mathbf{K}_{out} = 10^{-5}$ dimensionless permeability units) in the centre 
(Figure~\ref{fig:Porous_forward} (A)); the porosity is homogeneous in the whole domain and $\phi = 0.2$. The domain is 
initially saturated by the non-wetting phase with a saturation equal to ($1-S_{wirr}$). The wetting phase is injected 
(at a rate of $0.2$) over the left boundary displacing the non-wetting phase towards the right boundary; the other boundaries are closed to flow. The viscosity ratio of the phases is $1$. The time-step size is set to $0.025$ and the final time is set to $0.375$.

Figure~\ref{fig:Porous_forward} (B-D) shows the forward simulation of this test case. The flow goes around the low 
permeable inclusion.
The point of interest is located at the black diamond in Figure~\ref{fig:Porous_forward} (D) ($x=0.773$, $y = 0.5$, 
$z=0.59$), this location is chosen to force a flowpath that is not linear. 
The time of interest is at the end of time and therefore the functional $F$ 
is defined as the phase 1 saturation at the point and time of interest. 

\begin{figure}[h!]
  \begin{center}
    \includegraphics[width=0.80\textwidth]{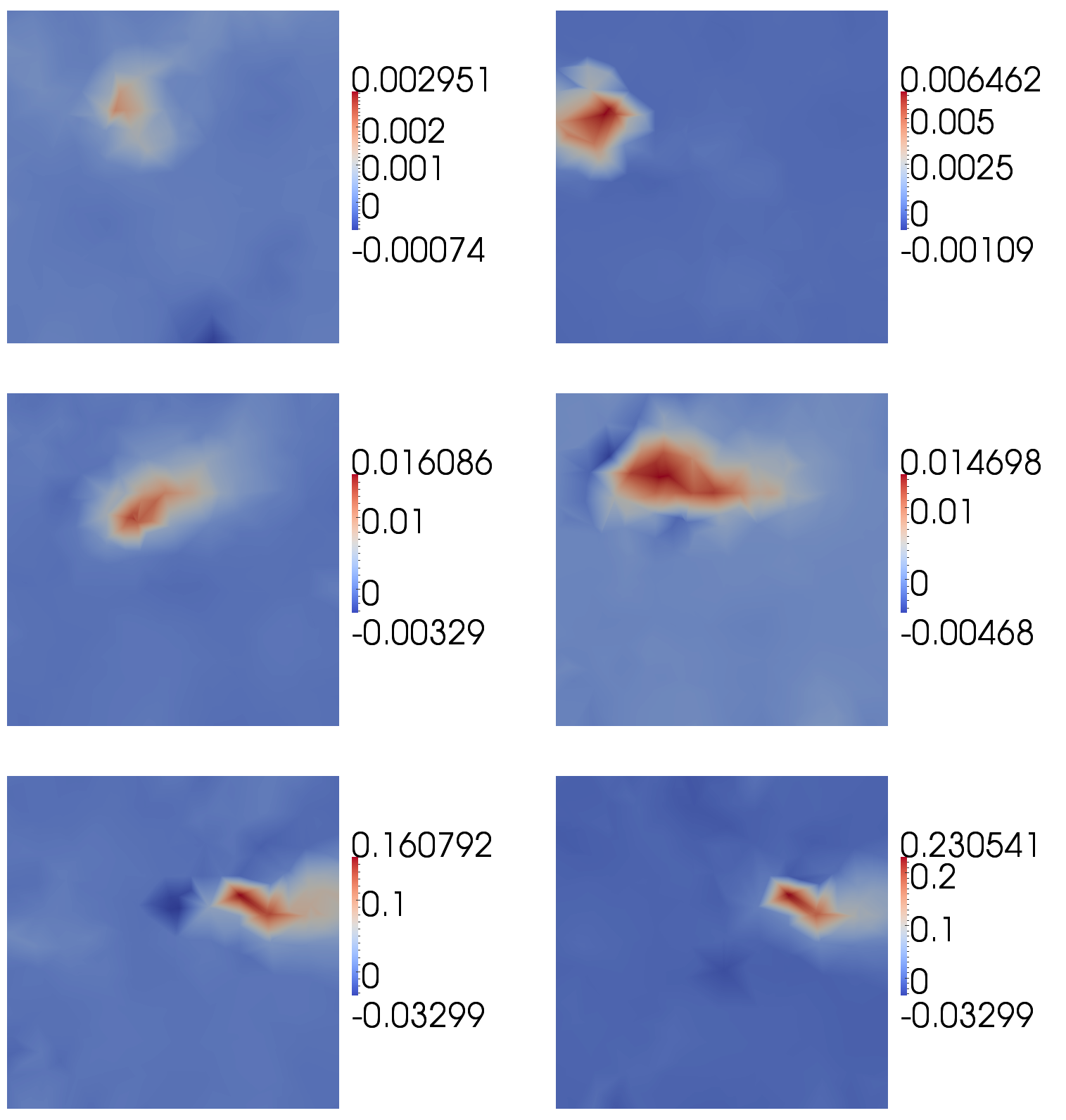}
    \caption{Results are generated by smoothing the perturbations, weighting them with the sensitivity map available at 
that time, orthogonalising and by using four time windows. Ensemble sizes of~20 (left) and~40 (right) are used. The 
sensitivity map is shown at times $t=0$ (top), $t=0.175$ (middle) and 
$t=0.375$ (bottom). \label{fig:Porous_backward}}
  \end{center}
\end{figure}

Two sets of experiments are performed, both with ensemble sizes of~20 and~40. One set is generated by performing 5 
smoothing iterations, weighting with the sensitivity map available at that time, orthogonalising and by using 4 time 
windows. The other set is generated without using these techniques. In the latter case, none of these results provided 
a meaningful sensitivity map, and therefore the results are not included. 
Figure~\ref{fig:Porous_backward} shows the sensitivity maps, the left column displays the results using an ensemble 
size of~20 and the right column, using an ensemble size of~40. It can be seen that, in this case, an ensemble size 
of~20 is not enough to provide a very reliable sensitivity map. Nonetheless, for an ensemble size of~40 the results are 
very accurate, as they show well-defined regions with little or no oscillation in the results.

\section{Conclusions}

We have proposed an optimised method for forming sensitivity maps based on ensembles. We have introduced
\begin{itemize}
 \item a new goal-based approach, which weights the perturbations with the most up-to-date sensitivities, thereby 
focusing the perturbations where they will most affect the goal, 
  \item time windows, which enable the perturbations to focus on 
developing sensitivities specifically for each time window and are thus more accurate,  
  \item re-orthogonalisation of the solution through time, which guarantees that a sensitivity map can be calculated 
and maximises the information that is obtained from each ensemble member.
\end{itemize}
\noindent
The number of ensemble members required to obtain sensitivity maps of a certain precision has 
been demonstrated to be greatly reduced by combining these techniques. In some of the cases presented here, just 10s of 
ensemble members are required, making this method extremely appealing when compared to formulating and implementing an 
adjoint model. Moreover, the presented method is generic and relies solely on perturbations obtained from the forward 
model. Thus the approach can be applied to forward models of arbitrary complexity, arising from areas such as coupled 
multi-physics, legacy codes or model chains, without the need to modify the code. There are a number of applications 
that could benefit greatly from this approach, including the computation of sensitivities for optimisation of sensor 
placement, optimisation for design or control, goal-based mesh adaptivity, assessment of goals (e.g. hazard assessment 
and mitigation in the natural environment), determining the worth of current data and data assimilation. 



\section*{Acknowledgments}
\noindent The authors are grateful for the support of the EPSRC through: 
the Smart-GeoWells Newton grant (P65437); 
Managing Air for Green Inner Cities (MAGIC, EP/N010221/1); the multi-phase flow programme grant 
(MEMPHIS, EP/K003976/1); multi-phase flow for subsea applications  
(MUFFINS, EP/P033148/1); and also funding from the European Union Seventh Frame work Programme (FP7/20072013) under 
grant agreement No.603663 for the research project Preparing for Extreme And Rare events in coastal regions (PEARL). 

\bibliographystyle{elsarticle-harv}
\bibliography{references} 

\end{document}